\documentclass[reprint,aps,prl,twocolumn,superscriptaddress,
floatfix,showpacs,amsmath,amssymb]{revtex4-1}

\usepackage{verbatim}
\usepackage{graphicx,color,bbm}% Include figure files
\usepackage{epsfig}
\usepackage{amsmath}
\usepackage{circuitikz}
\newcommand{\bra}[1]{\langle{#1} |}
\newcommand{\ket}[1]{|{#1}\rangle  }
\newcommand{\braket}[2]{\langle{#1}| {#2}\rangle}
\newcommand{\ketbra}[2]{\vert {#1} \rangle \langle{#2}\vert}

\newcommand{\tcred}[1]{\textcolor[rgb]{0.7,0,0}{#1}}

\usepackage[normalem]{ulem}
\begin{document}
\preprint{APS/123-QED}

\title{Coherent manipulation of noise-protected superconducting artificial 
atoms in the Lambda scheme}
\author{P.G. Di Stefano}
\affiliation{Dipartimento di Fisica e Astronomia,
Universit\`a di Catania, Via Santa Sofia 64, 95123 Catania, Italy.}
\affiliation{Centre for Theoretical Atomic, Molecular and Optical Physics,
School of Mathematics and Physics, Queen’s University Belfast, Belfast BT7 1NN, United Kingdom.}
\author{E. Paladino}
\affiliation{Dipartimento di Fisica e Astronomia,
Universit\`a di Catania, Via Santa Sofia 64, 95123 Catania, Italy.}
\affiliation{CNR-IMM  UOS Universit\`a (MATIS), 
Consiglio Nazionale delle Ricerche, Via Santa Sofia 64, 95123 Catania, Italy.}
\affiliation{Istituto Nazionale di Fisica Nucleare, Via Santa Sofia 64, 95123 Catania, Italy.}
\author{T.J. Pope}
\affiliation{Dipartimento di Fisica e Astronomia,
Universit\`a di Catania, Via Santa Sofia 64, 95123 Catania, Italy.}
\affiliation{School of Chemistry, Newcastle University, Newcastle NE1 7RU, United Kingdom}
\author{G. Falci}
\email[gfalci@dmfci.unict.it]{}
\affiliation{Dipartimento di Fisica e Astronomia,
Universit\`a di Catania, Via Santa Sofia 64, 95123 Catania, Italy.}
\affiliation{CNR-IMM  UOS Universit\`a (MATIS), 
Consiglio Nazionale delle Ricerche, Via Santa Sofia 64, 95123 Catania, Italy.}
\affiliation{Istituto Nazionale di Fisica Nucleare, Via Santa Sofia 64, 95123 Catania, Italy.}

\date{\today}
% 03.65.Yz Decoherence; open systems; quantum statistical methods
% 03.67.Lx Quantum computation architectures and implementations
% 03.67.Hk Quantum communication
% 89.70.+c Information theory and communication theory 
% 03.67.Pp Quantum error correction and other methods for protection against decoherence
% 42.50.Gy Effects of atomic coherence on propagation, absorption, and amplification of light; electromagnetically induced transparency and absorption
% 85.25.-j Superconducting devices
\begin{abstract}
We propose a new protocol for the manipulation of a three-level artificial atom in Lambda ($\Lambda$) configuration. It allows faithful, selective and robust population transfer analogous to stimulated Raman adiabatic passage ($\Lambda$-STIRAP), in last-generation superconducting artificial atoms, where protection from noise implies the absence of a direct pump coupling. It combines the use of a two-photon pump pulse with suitable advanced control, operated by a slow modulation of the phase of the external fields, leveraging on the stability of semiclassical microwave drives.
This protocol is a building block for manipulation of microwave photons in complex quantum architectures. Its demonstration would be a benchmark for the implementation of a class of multilevel advanced control procedures for quantum computation and microwave quantum 
photonics in systems based on artificial atoms.
\end{abstract}
% 03.67.Hk Quantum communication
% 03.67.-a Quantum information
% 03.67.Lx Quantum computation architectures and implementations
% 03.65.Yz Decoherence; open systems; quantum statistical methods
% 32.80.Qk
% 73.23.-b Electronic transport in mesoscopic systems
% 74.50. r Tunneling phenomena; point  contacts, weak links,         Josephson effects (for SQUIDs,         see 85.25.Dq; for Josephson         devices, see 85.25.Cp; for Josephson         junction arrays, see 74.81.Fa)
% 42.50.Gy Effects of atomic coherence on propagation, absorption, and amplification of light; electromagnetically induced transparency and absorption
% 85.25.-j Superconducting devices
\maketitle
%\section{Introduction}
Advanced control of multilevel quantum systems is a key requirement of quantum technologies~\citep{kb:210-nielsenchuang}, enabling tasks like multiqubit or multistate device processing~\cite{ka:211-timoney-nature-dressed,kr:211-younori-nature-multilevel,ka:209-lanyonwhite-natphys-quditprocessing} by adiabatic protocols, topologically protected computation~\cite{kb:209-pachos-topqcomp} or communication in distributed quantum networks\cite{ka:208-kimble-nature-qinternet,ka:214-darrigo-annals-hiddenent,ka:215:orieux-scirep-recoveryent}. 
These are currently investigated roadmaps towards the design of fault tolerant hardware, i.e. complex quantum architectures minimizing effects of decoherence~\cite{kr:210-ladd-nature-revqcomp,kr:213-devoretschoelkopf-science}. In this scenario artificial atoms (AAs) are very promising since, compared to their natural counterparts, they allow for a larger degree of integration~\cite{kr:208-schoelkopf-nature-wiring,ka:214-machaustinov-ncomms-supmetamaterials,ka:214-mohebbicory-japph-arraymicrostrip,ka:216-brechtschoelkopf-natqinfo}, on-chip tunability, stronger couplings~\cite{ka:210-niemczyksolano-natphys-ultrastrong} 
and easier production and detection of signals in the novel regime of microwave quantum photonics~\cite{ka:213-nakamurayam-ieee-microwphot}.  
Decoherence due to strong coupling to the solid-state environment~\cite{kr:214-paladino-rmp} is their major drawback. Over the years it has considerably softened~\citep{kr:213-devoretschoelkopf-science} {yielding last-generation superconducting 
devices with decoherence times in the range $\sim 1-100\,\mu\mathrm{s}$~\cite{
ka:211-bylander-natphys,ka:212-rigettisteffen-prb-trasmonshapphire,ka:214-sternsaclay-prl-fluxqubit3D}.}

\begin{figure}[t!]
\centering
\begin{minipage}[c]{0.35\columnwidth}
\includegraphics[width=\textwidth]{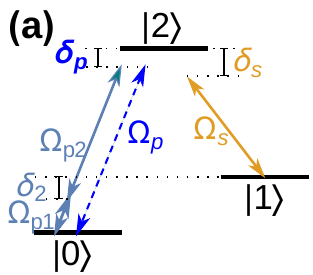}
\end{minipage}
\begin{minipage}[c]{0.55\columnwidth}
\includegraphics[width=\textwidth]{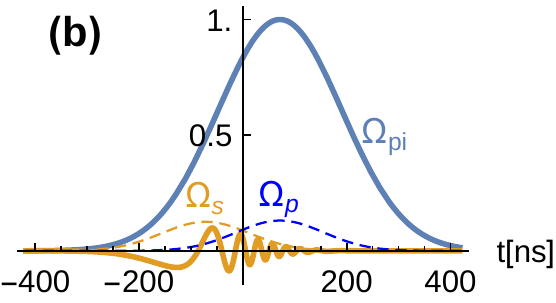}
\end{minipage}
\caption{(color online) (a) A three-level system with splittings
$\omega_{ij}:=E_j-E_i$ driven 
by two quasi resonant ac pump ($\Omega_p$) and Stokes ($\Omega_s$) fields, in the usual $\Lambda$ scheme
($\omega_{p} := \omega_{02} - \delta_p$, $\omega_{s}= \omega_{12} - \delta_s$). In 2+1 STIRAP 
the pump is operated by two pulses $\Omega_{pk}$   
at frequencies $\omega_{p1} := \omega_{01} - \delta_2$, $\omega_{p2}= \omega_{12}-\delta_p+\delta_2$
such that $\omega_{p1}+\omega_{p2}= \omega_{02}-\delta_p$.
(b) Pulses in conventional STIRAP (dashed lines) in the counterintuitive sequence, 
i.e. the Stokes pulse is shined before the pump pulse. Real part of the pulses in 2+1-STIRAP (solid lines): here $\Omega_{s} \text{e}^{i{\phi}_s(t)}$ shows the  
phase modulated control of Eq.(\ref{eq:chirps}). 
\label{fig:lambda-pulses-spectra}
}
\end{figure}

{Combining potential advantages of AAs is by no means straightforward. 
Protection from decoherence often implies strong constraints to available external control, 
which pose key challenges when larger architectures 
are considered~\cite{ka:216-brechtschoelkopf-natqinfo}. 
In this work we study a simple and paradigmatic example, namely a 
three-level AA driven by a two-tone electric field in the Lambda ($\Lambda$) configuration 
[Fig.\ref{fig:lambda-pulses-spectra}(a)]. 
Implementation of this control scheme in last-generation superconducting hardware may 
in principle benefit from low decoherence, which however is achieved at the expenses 
of suppressing the direct coupling of the pump field, and of possible limitations of 
selectivity in addressing specific transitions. In this work we show how to implement 
an efficient $\Lambda$ configuration in these conditions, and we propose a dynamical scheme 
allowing to operate quantum control. This solves the problem raised in the 
last decade by several theoretical proposals on the implementation of advanced control by a $\Lambda$-scheme in AAs~\cite{ka:205-liunori-prl-adiabaticpassage,ku:205-mariantoni-arXiv-microwfock,ka:206-siebrafalci-optcomm-stirap,ka:208-weinori-prl-stirapqcomp,ka:209-siebrafalci-prb,kr:211-younori-nature-multilevel}, which still awaits experimental demonstration.} 

{Quantum control via a dynamical $\Lambda$ scheme is very important because it may provide a fundamental building block for processing in complex architectures. Indeed adiabatic evolution 
may be used to trigger two-photon absorption-emission pumping cycles, which allow
for on demand manipulation of individual photons in distributed 
quantum networks, as proposed in the cavity-QED realm~\cite{ka:207-wilkrempe-science-singleatomph,kr:215-bergmannetal-jchemphys-revstirap}. 
Demonstrating control by a $\Lambda$ configuration in last-generation AAs  
would extend this scenario to the microwave arena, opening the 
perspective of performing demanding protocols in highly integrated solid-state quantum architectures~\cite{ka:214-mohebbicory-japph-arraymicrostrip,ka:216-brechtschoelkopf-natqinfo}, 
which are usually subject to specific design constraints~\cite{ka:215-distefano}. Examples  
are adiabatic holonomic quantum computation~\cite{ka:208-dmoller-prl-adiabgates}, information transfer and entanglement generation~\cite{ka:208-weinori-prl-stirapqcomp,ka:204-yang-prl-fluxstirap,ka:204-kis-prb-fluxstirap} between remote nodes, and other sophisticated 
control protocols~\cite{kr:207-kral-rmp-controladpass}.}

The $\Lambda$ scheme is described by the standard Hamiltonian in the rotating-wave 
approximation (RWA), which 
in a double rotating frame reads~\cite{kr:201-vitanov-advatmolopt}
\begin{equation}
\label{eq:stirap-H}
H = \frac{\Omega_p}{2} \ketbra{0}{2}+\frac{\Omega_s}{2} \ketbra{1}{2}+ \mbox{h.c.}+\delta\ketbra{1}{1}+\delta_p\ketbra{2}{2}
\;.
\end{equation}
Here  $\Omega_{p,s}$ are the Rabi frequencies of the (pump and Stokes, respectively) external fields, with single-photon detunings $\delta_{p,s}$, and $\delta=\delta_p -\delta_s$ is the  two-photon detuning [Fig.\ref{fig:lambda-pulses-spectra}(a)]. 
This latter is a key parameter, since for $\delta = 0$ 
the system admits an exact dark state $\ket{D}:=(\Omega_p^2+\Omega_s^2)^{-\frac{1}{2}} (\Omega_s\ket{0}-\Omega_p\ket{1})$: the system is trapped in $\ket{D}$ 
despite the two fields triggering transitions towards $\ket{2}$, a 
striking destructive quantum interference phenomenon named coherent population trapping~\citep{kr:201-vitanov-advatmolopt,ka:210-kellypappas-prl-cpt}. Sensitivity to $\delta$ is critical since no exact dark state exists for $\delta \neq 0$, even if partial trapping is still 
supported. Controlling the dynamics of $\ket{D}$ would lead to the observation of basic interference effects, allowing also important applications.
A representative example is stimulated Raman adiabatic passage (STIRAP)~\citep{kr:198-bergmann-rmp-stirap,kr:201-vitanov-advatmolopt}, a powerful technique allowing faithful and selective population transfer in atomic physics. By adiabatically varying $\Omega_{p,s}(t)$ 
% in the Hamiltonian Eq.(\ref{eq:stirap-H}) 
in the so called {\em counterintuitive sequence} [Fig.\ref{fig:lambda-pulses-spectra}(b)],
the state  $\ket{D(t)}$ evolves from $\ket{0}$ to $\ket{1}$, in the absence of a direct coupling and never populating the intermediate state $\ket{2}$. STIRAP is a benchmark for multilevel advanced control. Its robustness against imperfections and disorder
may allow to develop new protocols~\cite{kr:207-kral-rmp-controladpass,ka:211-timoney-nature-dressed} with important applications in hybrid networks, composed of many AAs or microscopic spins interacting with quantized modes~\citep{kr:213-xiangnori-rmp-hybrid}. 

Several works in the last decade~\cite{ka:205-liunori-prl-adiabaticpassage,ku:205-mariantoni-arXiv-microwfock,ka:206-siebrafalci-optcomm-stirap,ka:208-weinori-prl-stirapqcomp,ka:209-siebrafalci-prb,kr:211-younori-nature-multilevel} proposed implementations of $\Lambda$-STIRAP 
in superconducting AAs. However dynamics in the $\Lambda$ scheme has not yet been experimentally demonstrated for a fundamental reason: protection against low-frequency noise~\cite{kr:214-paladino-rmp}
necessary to achieve large decoherence times is attained by operating the device by a Hamiltonian with exact or approximate symmetries~\cite{ka:202-vion-science,ka:205-falci-prl}. In 
particular low decoherence in last-generation superconducting AAs is obtained 
by enforcing parity symmetry, which however implies cancellation of the coupling to the pump field~\cite{ka:205-liunori-prl-adiabaticpassage,ka:206-siebrafalci-optcomm-stirap,ka:209-siebrafalci-prb,kr:211-younori-nature-multilevel,note:suppl-mat}. $\Lambda$-STIRAP
could be observed by breaking the symmetry of the device~\cite{ka:205-liunori-prl-adiabaticpassage,ka:206-siebrafalci-optcomm-stirap}, but at the expenses of an increased noise level. Analysis of a case study~\cite{ka:213-falci-prb-stirapcpb} has shown that efficiency, 
{i.e. the final population of the target state $\ket{1}$},  
does not exceed $\gtrsim 70\%$. 

In order to design an effective $\Lambda$ scheme, i.e. allowing efficient coupling 
at symmetry, where decoherence times are large, we first replace the direct pump pulse 
by a two-photon process,
which yields overall the "2+1" $\Lambda$ scheme [see Fig.\ref{fig:lambda-pulses-spectra}(a)]. This configuration is however known to lack robustness against fluctuations of the parameters~\citep{ka:198-yatsenko-pra-stirap21th1,ka:198-guerin-pra-stirap21staticcomp,ka:198-bohmer-pra-stirap21exp,kr:215-bergmannetal-jchemphys-revstirap}. To overcome this problem
we supplement the "2+1" $\Lambda$ scheme by suitable advanced control, which turns out to be
the key ingredient for achieving both $\sim 100 \%$ population transfer efficiency and robustness. We address two classes of last-generation AAs, based on the 
"flux-qubit"~\cite{ka:200-walmooij-science-superposition,ka:211-bylander-natphys} and on the "transmon"~\cite{ka:204-wallraff-superqubit,ka:207-koch-pra-transmon,ka:212-rigettisteffen-prb-trasmonshapphire} design, respectively. 

%\section{STIRAP 2+1}
We start our analysis from 
the full Hamiltonian
\begin{equation}
\label{eq:driven-H}
H:=H_0 + H_C(t)  
\end{equation}
where $H_0 := \sum_j E_j \ketbra{j}{j}$ models the 
undriven AA. The control 
$H_C = {\cal Q} \,{\cal A}(t)$ is operated by a three-tone field 
${\cal A}(t)= \sum_{m=p1,p2,s}{\cal A}_{m}(t) 
\cos [\omega_m t- {\phi}_{m}(t)]$. It is coupled to the operator ${\cal Q}$, 
corresponding to the electric dipole for natural atoms. In AAs it     
is, for instance, the charge operator in the transmon~\cite{ka:207-koch-pra-transmon,ka:212-rigettisteffen-prb-trasmonshapphire} 
or the loop current in the flux qudit~\cite{ka:200-walmooij-science-superposition,ka:211-bylander-natphys}. Symmetries in the  
Hamiltonian $H_0$ imply that matrix elements ${\cal Q}_{ii}={\cal Q}_{02}= 0$. 
External fields have suitable carrier frequencies (see Fig.\ref{fig:lambda-pulses-spectra}(a))
% $\omega_{p1}:=\omega_{01}-\delta_2$, $\omega_{p2} := \omega_{12} +\delta_2-\delta_p$ and $\omega_s = \omega_{12}-\delta_s$.
and a slowly varying 
modulation of the phases ${\phi}_m(t)$, for $m=s,p1$. 
Rabi angular frequencies are defined as $\Omega_{p1}(t):={\cal Q}_{01} {\cal A}_{p1}(t)$, $\Omega_{p2}(t):={\cal Q}_{12} {\cal A}_{p2}(t)$, $\Omega_{s}(t):={\cal Q}_{12} {\cal A}_{s}(t)$. For simplicity 
we take $\delta_p=\delta_s=0$ and equal peak amplitudes $\Omega_r$ for both the $\Omega_{pk}(t)$, where $k=1,2$, 
considering Gaussian pulses
\begin{equation}
\label{eq:pulses}
\Omega_s(t) = \Omega_0\, \text{e}^{-\left( \frac{t+\tau}{T} \right)^2} ; \qquad
\Omega_{pk}(t) = \Omega_r\, \text{e}^{-\frac{1}{2}\left( \frac{t-\tau}{T} \right)^2}
\end{equation}
We use the delay~\cite{kr:201-vitanov-advatmolopt} $\tau = 0.6 \,T >0$ which implements the counterintuitive sequence. 

Our goal is to reproduce Eq.(\ref{eq:stirap-H}) as an effective Hamiltonian yielding STIRAP, by properly 
shaping the control $\{{\phi}_m(t)\}$.  
We first consider an AA with a highly anharmonic spectrum, $\omega_{12} \gg \omega_{10}$,  where each transition can be selectively addressed. Therefore we can safely
neglect the strongly off-resonant ${\cal A}_{p1}$ (${\cal A}_{p2}$ and ${\cal A}_{s}$) in $\braket{1|H}{2}$ ($\braket{0|H}{1}$), and also perform the RWA. 
The three-level Hamiltonian in the interaction picture reads
\begin{equation}
\label{eq:2+1-3ls-H}
\begin{aligned}
H_3 = & \frac{1}{2}
\big\{\Omega_{p1}(t)\, \text{e}^{-i\delta_2 t} \ketbra{0}{1} + \big[ \Omega_{p2}(t) \, 
\text{e}^{i[\delta_2 t -{\phi}_{p2}(t)]}+
\\&+ 
\Omega_{s}(t) \, \text{e}^{-i{\phi}_{s}(t)} 
\big] \ketbra{1}{2}  \big\} + \mbox{h.c.}
\end{aligned}
\end{equation}
If the two pump pulses are strongly dispersive, 
$|\delta_2|/\Omega_r\gg1$, they 
implement an effective two-photon $\ket{0} \leftrightarrow \ket{2}$ pulse which does not  
populate $\ket{1}$~\cite{kr:201-vitanov-advatmolopt}. In this regime we derive an effective Hamiltonian 
from the Magnus expansion of time-evolution operator corresponding to $H_3$ ~\cite{note:suppl-mat,ka:168-waugh-prl-average-hamiltonian}.
%Eq.(\ref{eq:2+1-3ls-H}) 
The relevant contributions are found up to second order, which 
captures the coarse-grained dynamics averaged over a convenient time scale $\Delta t$ such that $\Delta t \,|\delta_2| \gg 1$ but $\Delta t \,\Omega_r,\Delta t/T, \Delta t\, 
|\dot{{\phi}}_i(t)| \ll 1$\cite{note:suppl-mat}. Then in the same rotating frame of Eq.(\ref{eq:stirap-H}) we obtain
\begin{equation}
\label{eq:2+1-3ls-ave-H}
\begin{aligned}
H_{eff} =& 
[(\dot{{\phi}}_{p2}-\dot{{\phi}}_{s})-(S_2+2S_1)]
\ketbra{1}{1}+\\ &+
(\dot{{\phi}}_{p2}+S_2-S_1)\ketbra{2}{2} +
\\ &+
\frac{1}{2} \big[ 
\big(\Omega_p \ketbra{0}{2}+ \Omega_s \ketbra{1}{2} \big)
+ \mbox{h.c.} \big]
\end{aligned}
\end{equation}
where $\Omega_p(t) = - {\Omega_{p1}\Omega_{p2}}/(2\delta_2)$ is 
the two-photon effective pump field and $S_k(t) = - {\Omega_{pk}^2}/(4\delta_2)$ are dynamical Stark shifts. 
We see that if we define 
$\delta_p := \dot{{\phi}}_{p2}+S_2-S_1$ and $\delta := \dot{{\phi}}_{p2}-\dot{{\phi}}_{s}-(S_2+2S_1)$, 
Eq.(\ref{eq:2+1-3ls-ave-H}) reproduces Eq.(\ref{eq:stirap-H}) 
identifying the effective $\Lambda$ system.   

\begin{figure*}
\includegraphics[width=0.8\columnwidth]{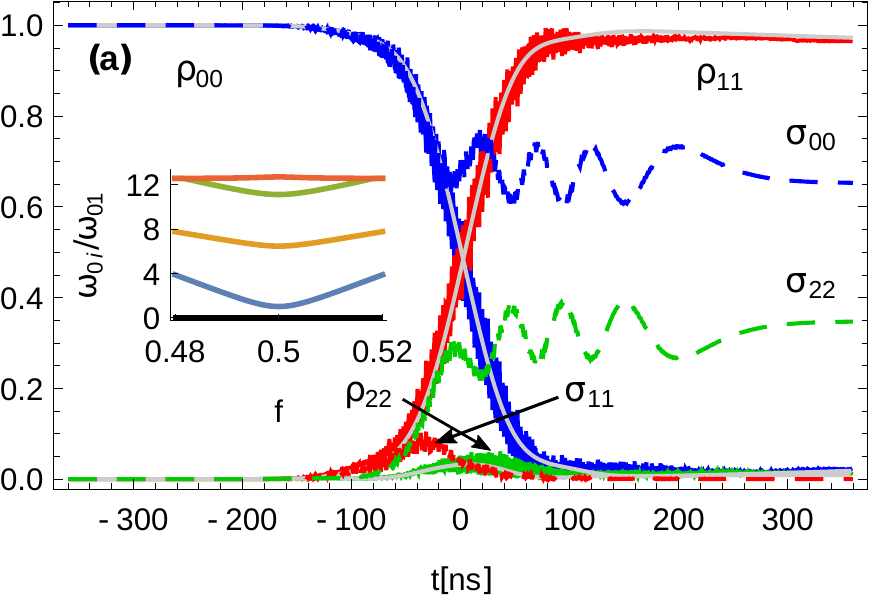}
\qquad
\includegraphics[width=0.8\columnwidth]{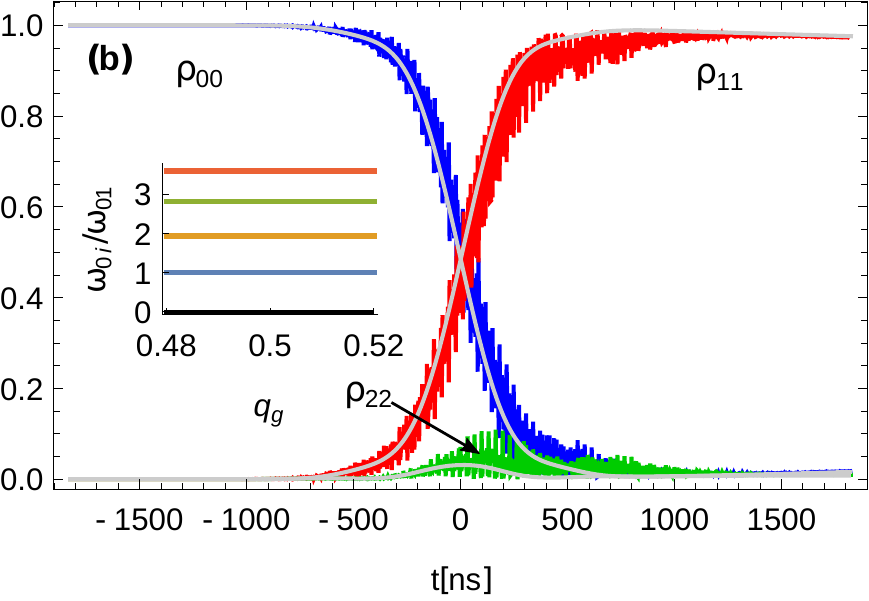}
\caption{(color online) Population histories $\rho_{00}(t)$ (blue) $\rho_{11}(t)$ (red) and $\rho_{22}(t)$ (green). (a) For a flux AA biased at the symmetry point $f=1/2$ (spectrum in the inset), with the phase modulation Eq.\ref{eq:chirps}. We used $\Omega_r/2\pi=200\text{MHz}$ and $\delta_2 =-5 \Omega_r$, for the 
two-photon pump, yielding $\Omega_0 = |\Omega_r/2\delta_2| = 20 \text{MHz}$. Good adiabaticity,
$\Omega_0 T=15$, is obtained with $T =0.12 \,\mu\mathrm{s}$ and $\tau = 0.6 \,T$. Results refer to the device Ref.~\onlinecite{ka:211-bylander-natphys} and account for leakage and effects of noise.
In the absence of phase modulation, population histories $\sigma_{ii}$ in the absence of decoherence (dashed lines) show no population transfer. 
(b) Same results for a transmon (spectrum in the inset) with phase modulation Eq.~(\ref{eq:new-chirps}). 
Here $\Omega_0 = 3.9 \,\text{MHz}$ from Eq.(\ref{eq:W0}), $T=0.6 \,\mu\mathrm{s}$ and 
$\tau = 0.6 \,T$.
For both designs the approximate effective 
dynamics (gray thin lines above the exact population histories), obtained respectively from Eq.(\ref{eq:2+1-3ls-ave-H}) and Eq.(\ref{eq:ave-H-correction}),
reproduces remarkably well the coarse grained time-evolution.
\label{fig:populations}}
\end{figure*}
We now look for external control yielding STIRAP. It is convenient to take equal pulse 
amplitudes in $H_{eff}$, thereby $\Omega_0 =\Omega_r^2/(2|\delta_2|)$, and the necessary 
condition for adiabaticity~\cite{kr:198-bergmann-rmp-stirap}  
sets the time scale $T > 10/\Omega_0$. We finally adjust the system at both single and 
two-photon effective resonance by choosing the phase modulation according to
\begin{equation}
\label{eq:chirps}
\begin{aligned}
\dot{{\phi}}_{p2} =& S_1-S_2 \quad ; \quad
\dot{{\phi}}_{s}=&- (S_1 + 2 S_2)
\end{aligned}
\end{equation}
This is a key point of our analysis: performing the latter step is crucial since STIRAP would fail otherwise 
[see Fig.\ref{fig:populations}(a), dashed lines]. Indeed the dynamical Stark shifts 
$S_k(t)$ are of the same order of the effective coupling $\Omega_p(t)$. Therefore if uncompensated they would determine large stray detunings, in particular $\delta(t) = - (S_2+2S_1)$ would destroy the dark state. 

The phase modulation in Eq.(\ref{eq:chirps}) is obtained in closed form as a function of the pulse envelopes ${\cal A}_m(t)$ 
by a simple integration. Inserted in the control of the full Hamiltonian Eq.(\ref{eq:driven-H}) it yields the goal we set, 
namely $\sim 100\%$ efficiency is recovered [see Fig.\ref{fig:populations}(a), solid lines]. 

An important point is that solutions ${\phi}_m(t)$ of interest are slowly varying,  
consistent with our assumption. This is also clear from in Fig.\ref{fig:lambda-pulses-spectra}(b), where the modulation of the Stokes pulse for equal $\Omega_{pi}$s, i.e. $S_1=S_2$, is shown. It is worth stressing 
the remarkable agreement between the full dynamics and the approximation by $H_{eff}$ 
(gray lines in Fig.\ref{fig:populations}), which we will use later to estimate appropriate figures for
$\Omega_r$, $\Omega_0$ and $T$.

{Noise sources coupled via the operator ${\cal Q}$ are usually the most detrimental for decoherence. Effects of low-frequency noise from this "port" can be suppressed by 
designing a Hamiltonian with suitable symmetries, a strategy that has yielded very large decoherence times in last-generation superconducting qubits. On the other hand high-frequency fluctuations 
from the ${\cal Q}$-port are the relevant sources of quantum Markovian noise. 
Pure dephasing is due to residual non-Markovian noise from sources coupled to operators 
orthogonal to ${\cal Q}$.}
The impact of noise is studied 
using a phenomenological picture~\cite{ka:205-falci-prl,kr:214-paladino-rmp,note:suppl-mat},  
accounting for both Markovian and non-Markovian relevant noise sources. 
Markovian quantum noise is described by a "dissipator" ${\mathcal L}_D$ 
in a Master Equation of the Lindblad form 
\begin{equation}
\label{eq:lowfreq}
\dot{\rho}(t|\mathbf{\tilde{x}})
=-i[H(\mathbf{\tilde{x}}(t)),\rho(t|\mathbf{\tilde{x}})] + {\mathcal L}_D \rho(t|\mathbf{\tilde{x}})
\end{equation}
whose solution has to be averaged over a stochastic process $\mathbf{\tilde{x}}(t)$ describing 
individual realizations of the non-Markovian classical noise. 
For noise with $\sim 1/f^\alpha$ low-frequency spectrum the leading effect 
is captured by retaining only the contribution of quasistatic stray bias $ \mathbf{\tilde{x}}(t) \to \mathbf{\tilde{x}}$ of the artificial atom~\cite{ka:205-falci-prl,kr:214-paladino-rmp}, with a suitable 
Gaussian distribution. In this picture stray bias determine
fluctuations of energies $\Delta E_i$ and of matrix elements $\Delta {\cal Q}_{ij}$, which 
translate respectively in fluctuations of the detunings $\tilde{\delta}=\Delta (E_1-E_0)$ and $\tilde{\delta}_p =\Delta (E_2-E_0)$ and of the Rabi frequency $\tilde{\Omega}_0$. 
Only the former turn out to be 
important~\cite{kr:201-vitanov-advatmolopt,ka:213-falci-prb-stirapcpb}, thereby Eq.(\ref{eq:lowfreq}) reduces to the structure $\dot{\rho}=i[\rho,H(\delta,\delta_p)] + {\mathcal L}_D \rho$, where 
detunings undergo correlated fluctuations 
$(\tilde{\delta},\tilde{\delta}_p)$ 
induced by $\mathbf{\tilde{x}}$, the full dynamics emerging from proper averaging. 
%$\rho(t|\delta,\delta_p)$ over correlated fluctuations of detunings.

In practical cases a single additional port must be considered, with associated stray bias $\tilde{x}$. Then fluctuations have a simple linear correlation $\tilde{\delta}_p(\tilde{x}) = a \,\tilde{\delta}(\tilde{x})$, where $a$ is determined by the parametric dependence of the spectrum on $\tilde{x}$ (See Ref. \onlinecite{note:suppl-mat}). In this case experiments characterizing the {\em qubit} dynamics
yield all the needed statistical properties 
of $(\tilde{\delta},\tilde{\delta}_p)$, since the standard deviation of $\tilde{\delta}$ is $\sigma_\delta=\sqrt{2}/T_2^\prime$, where $1/T_2^\prime:=1/T_2^*-1/(2T_1)$ is the qubit non-Markovian pure dephasing rate~\cite{kr:214-paladino-rmp} and $T_1$ the qubit relaxation time. The multilevel dynamics is 
obtained by averaging over a Gaussian distribution, $p(\tilde{\delta})=(2\pi \sigma_{\delta}^2)^{-1/2} \text{e}^{-\tilde{\delta}^2/({2\sigma_{\delta}^2})}$, the solution 
$\rho(t|\tilde{\delta},a\tilde{\delta})$ of 
Eq.(\ref{eq:lowfreq}). We use  
the Markovian dissipator 
\begin{equation}\label{eq:ME}
\begin{aligned}
{\mathcal L}_D \rho =&
- \frac{1}{2T_1}([\ketbra{1}{1},\rho]-2 \ketbra{0}{1}\rho\ketbra{1}{0}) +\\ &- \frac{k}{2 T_1}([\ketbra{2}{2},\rho]-2 \ketbra{1}{2}\rho \ketbra{2}{1})
\end{aligned}
\end{equation}
accounting for the two allowed transitions in the lowest three levels. 
We assume that ${\mathcal L}_D$ does not depend explicitly on $\mathbf{\tilde{x}}$,
and we retain only spontaneous decay, which is the only relevant process 
at low enough temperature~\cite{ka:213-falci-prb-stirapcpb}. The constant $k \simeq [Q_{21}/Q_{10}]^2\, \,S(E_2-E_1)/S(E_1-E_0)$ depends essentially on the design of the device and, in a much weaker way, on the power spectrum $S(\omega)$, which is often ohmic at the relevant frequencies~\cite{kr:214-paladino-rmp,note:suppl-mat}.

%Modulated phase control could be successfully applied to HNP flux-qudits. 
In  Fig.~\ref{fig:populations}(a) we present results for the 
four-junctions SQUID of Ref.\cite{ka:211-bylander-natphys}. They show 
that $\Lambda$-STIRAP with $\sim 100\%$ efficiency
is obtained using $T \simeq 0.12 \,\mathrm{\mu s}$.  
We simulate the dynamics for the lowest six states of the full device Hamiltonian 
$H_0$\cite{note:suppl-mat}, verifying that leakage from the three-level subspace is negligible ($\sum_{j\ge 3}\rho_{jj} < 2 \times 10^{-4}$). For this device relaxation
($T_1 = 12 \,\mathrm{\mu s}$~\cite{ka:211-bylander-natphys}) and the associated Markovian dephasing 
are due to flux noise, whereas critical current and charge noise determine 
non-Markovian fluctuations, yielding the overall $T_2^* = 2.5\,\mu\mathrm{s}$. 
We find the remarkable $\gtrsim 97\%$ efficiency, which is essentially limited  by $T_1$ only. %{\sout{LO LEVIAMO? The $\ket{2}\to\ket{1}$ decay process is weighted by $k=0.73$ and determines a $\sim 0.5\%$ increase of the efficiency.}}

%\section{STIRAP in transmon}
We now turn to AAs  based on the transmon design~\cite{ka:204-wallraff-superqubit,ka:207-koch-pra-transmon,note:suppl-mat}. 
Successful implementation of $\Lambda$-STIRAP in this class of devices would be very important, since they display the largest decoherence times observed so far~\cite{ka:212-rigettisteffen-prb-trasmonshapphire,kr:213-devoretschoelkopf-science}, and  
% the dipole moment is well approximated by a ladder climbing operator $\hat{{\cal Q}}:=\sum_j {\cal Q}_{jj+1} \ketbra{j}{j+1}+h.c.$ [Kock]. 
% Other matrix elements vanish or are  very small, besides corresponding to transitions strongly detuned with respect to the control field. 
offer the perspective of fabricating highly integrated architectures~\cite{ka:214-mohebbicory-japph-arraymicrostrip,ka:216-brechtschoelkopf-natqinfo}, with a rich arena of applications. 
These AAs have a nearly harmonic spectrum [inset of Fig.\ref{eq:2+1-3ls-H}(b)], quantified by $\alpha := \omega_{12} - \omega_{01}$ and 
$\beta := \omega_{23} - \omega_{12}$ for the four lowest energy levels.
Values of $|\alpha| \simeq |\beta| \lesssim \omega_{01}/10$ ensure very large decoherence times, at the expenses however of limiting selectivity in addressing the desired transitions with strong fields. Harmonicity is a severe drawback for operating STIRAP and indeed the protocol outlined for flux-based AAs would fail in the transmon. 
In order to find the proper effective Hamiltonian we must: (a) include selected off-resonant terms of the control, relaxing the 
quasi-resonant approximation; (b) consider explicitly a fourth level $\ket{3}$ since it will determine 
Stark shifts which must be accounted for. We neglect the coupling to the cavity used in the transmon as a measuring apparatus and at this stage we also assume the RWA,
%{(see Fig. in figurea mettere che i Bloch Siegert sono comresi ma non importanti)}
% therefore and will be assumed for the sake of simplicity.  
so we consider the  Hamiltonian $H = H_3 + \tilde{H}$ in the interaction picture, with extra terms 
\begin{equation}
\label{eq:RWA-H-CPB}
\begin{aligned}
\tilde{H} =& \Big\{
\frac{1 }{2} {\cal A}_{p1} \text{e}^{- i (\delta_2+\alpha) t }\big[  
{\cal Q}_{12}  \ketbra{1}{2} +
{\cal Q}_{23} \text{e}^{-i \beta t} \ketbra{2}{3} \big]
\\+&
\frac{1}{2} {\cal A}_{p2} \,\text{e}^{i \delta_2 t }
\big[{\cal Q}_{01} \text{e}^{i[ \alpha t + {\phi}_{p2}(t)]}  \ketbra{0}{1} 
\\+& 
{\cal Q}_{23} \,\text{e}^{-i[\beta t - {\phi}_{p2}(t)]} \ketbra{2}{3} \big] + 
\mbox{h.c.} \Big\}
\end{aligned}
\end{equation}
The stray $\tilde{H}$ produces non negligible effects due to the fact that
anharmonicities $|\alpha|, |\beta|$ are small and large ${\cal A}_{pk}$ are needed to yield 
a sufficient effective dispersive pump drive. 
% $ \sim \Omega_r^2/\delta_2$ while keeping the dispersive regime. 
Since  ${\cal A}_{s}$ needs not to be large, the corresponding terms can be neglected. 
A convenient choice of parameters turns out to be $|\delta_2| \gtrsim |\alpha|, |\beta|$. 
In this regime we obtain the following three-level effective Hamiltonian in the 
rotating frame 
\begin{equation}
\label{eq:ave-H-correction}
\begin{aligned}
&H_{eff} = \Big(\frac{\Omega_p}{2}
\,\ketbra{0}{2} 
+ \frac{\Omega_s}{2} \,\ketbra{1}{2}  + 
\mbox{h.c.} \Big) \\
&\quad+  \sum_{k,i\neq j}S_{ji}^k \ketbra{i}{i} + \dot{{\phi}}_{p2} \ketbra{1}{1}+
(\dot{{\phi}}_{p2}-\dot{{\phi}}_{s}) \ketbra{2}{2}
\end{aligned}
\end{equation}
where the effective pump coupling is now
\begin{equation}
\label{eq:W0}
\Omega_p = -\frac{\Omega_{p1}\Omega_{p2}}{4 \delta_2} \frac{\alpha}{\alpha + \delta_2} 
\end{equation}
and the dynamical Stark shifts of level $j$ due to the coupling to level $i$ under the action of the $pk$ field is given by
\begin{equation}
\label{eq:2ph-pump-stark-shifts}
S_{ij}^k(t):=\left\lvert \frac{{\cal A}_{pk}(t){\cal Q}_{ij}}{2} \right\rvert^2 \left( \frac{1}{\omega_{ij}-\omega_{pk}} +
\frac{1}{\omega_{ij}+\omega_{pk}} \right)
\end{equation}
This antisymmetric form for $i \leftrightarrow j$ accounts also for Bloch-Siegert shifts, which 
are however small in all cases treated in this work. 
Notice that Eq.(\ref{eq:ave-H-correction}) includes three levels since  
levels $i > 2$ only yield Stark shifts. We again let $\delta_p=\delta=0$, 
thereby 
in order to obtain large STIRAP 
efficiency we  modulate phases according to
\begin{equation}
\label{eq:new-chirps}
\dot{{\phi}}_{p2} = \sum_{k,j} (S_{j0}^k-S_{j2}^k) \; ;\quad
\dot{{\phi}}_{s}= \sum_{k,j} (S_{j1}^k-S_{j2}^k)
\end{equation}
If we now use this modulation in the full Hamiltonian 
Eq.(\ref{eq:driven-H}) we recover $\sim 100\%$ efficiency
[see Fig.~\ref{fig:populations}(b)]. 
Again the full dynamics is remarkably well approximated by the Magnus expansion. 
Results refer to the transmon of Ref.\onlinecite{ka:212-rigettisteffen-prb-trasmonshapphire} and account for both leakage and effects of noise. Coherence is again essentially limited by $T_1 = 70 \,\mathrm{\mu s}$, thereby noise has negligible effects, also allowing for multiple STIRAP-like cycles. %{\sout{LO LEVIAMO? The $\ket{2}\to\ket{1}$ decay process is weighted by $k=2$ in this case and determines a $\sim 1\%$ increase of the efficiency.}}

\begin{figure}[t]
\begin{minipage}{0.48 \columnwidth}
\includegraphics[width=\columnwidth]{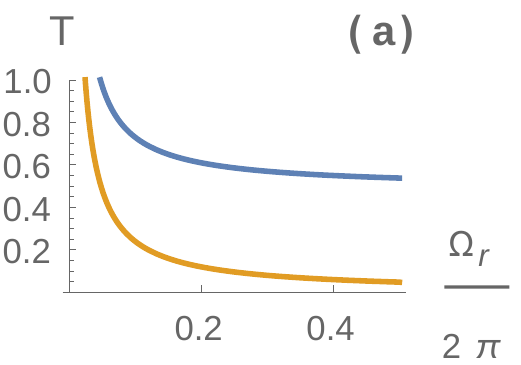}
\includegraphics[width=\columnwidth]{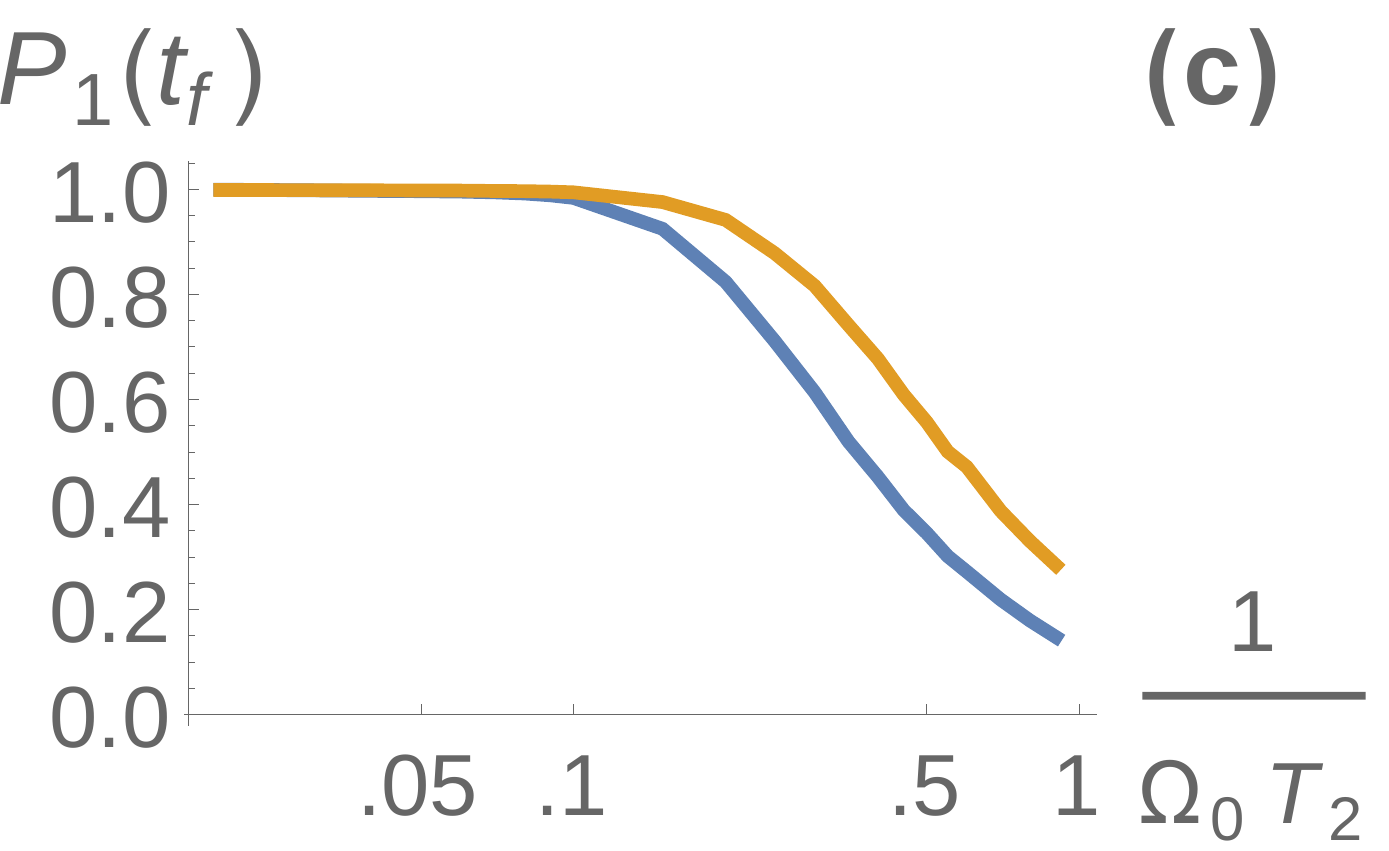}
\end{minipage}
\begin{minipage}{0.48 \columnwidth}
\includegraphics[%height=50mm,
width=\columnwidth]{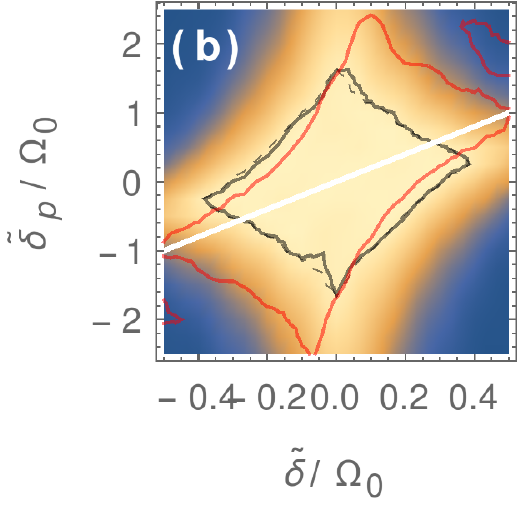}
\end{minipage}
\caption{(color online) Parametric robustness of the protocol. {(a) Effective duration of the protocol $T=15/\Omega_0$ ($\mathrm{\mu s}$ units) vs $\Omega_r/2\pi$ ($\mathrm{GHz}$ units) for the transmon of Ref.~\onlinecite{ka:212-rigettisteffen-prb-trasmonshapphire} (Eq.(\ref{eq:W0}), upper curve) and 
for the  flux-qudit of Ref.\onlinecite{ka:211-bylander-natphys} (limit $\alpha \gg \delta_2$, lower curve), at fixed $|\delta_2|/\Omega_r=5$. 
%\sout{The protocol for the transmon is always slower due to the nearly-harmonic spectrum [see Eq.(\ref{eq:W0})]}. 
(b) Efficiency versus stray detunings $\tilde{\delta}, \tilde{\delta}_p$ for the transmon 
design, showing the robustness of the protocol Eq.(\ref{eq:new-chirps}). 
The solid black inner curve encloses the region of efficiency $> 95\%$; results of the approximation $H_{eff}$, Eq.(\ref{eq:ave-H-correction}) are also reported (dashed curve) which show again the remarkable accuracy of the effective theory. The white straight line $\tilde{\delta}_p = 2 \tilde{\delta}$ represents the correlated quasitatic fluctuations of the stray detunings. The red outer curve encloses the $> 95\%$ efficiency area for $r=2$: it is seen that robustness along the line further increases using $r> 1$.
(c) Efficiency vs $T_2$ for the transmon, $r=1$ (lower line) and $r=2$ (upper line). 
 }
\label{fig:diamond}
}
\end{figure}

Notice that Eq.(\ref{eq:W0}) implies that the effective peak $\Omega_p$ 
saturates to the value $-\alpha/2(\delta_2/\Omega_r)^2$,  
for increasing $\Omega_r$ at constant $\delta_2/\Omega_r \gg 1$
[see Fig.\ref{fig:diamond}(a)]. For this reason the duration of the protocol for the transmon 
[$T=0.6 \,\mathrm{\mu s}$ in Fig.\ref{fig:populations}(b)] is larger than for the flux-based AA. 
More generally, shining larger external fields 
to shorten the protocol is useful only to some extent [see Fig.\ref{fig:diamond}(a)], but in 
devices with the largest coherence times this is not a limitation.

%\section{decoerenza}
{Robustness of the protocol is a crucial issue, since the success of conventional 
STIRAP lies in the striking insensitivity to small variations of control parameters.
In the early proposal of "2+1" $\Lambda-$STIRAP, lack of efficiency due to the 
stray dynamical Stark-shift was cured by using fields with a small static two-photon detuning~\cite{ka:198-yatsenko-pra-stirap21th1,ka:198-guerin-pra-stirap21staticcomp,ka:198-bohmer-pra-stirap21exp}, but unfortunately the resulting protocol was not robust~\cite{kr:215-bergmannetal-jchemphys-revstirap}. 
{Instead our control scheme is tailored to guarantee the same robustness of conventional STIRAP}. In 
Fig.\ref{fig:diamond}(b) we show sensitivity against fluctuations of the detunings  
of phase modulated STIRAP in the trasmon, which is potentially the most unfavourable case. 
For the example shown, frequency fluctuations of the microwave fields $\lesssim 1\,\textrm{MHz}$ 
still guarantee $\gtrsim 95\,\%$ efficiency. This important result would be hardly attainable for natural atoms driven at optical frequencies~\cite{ka:198-yatsenko-pra-stirap21th1,ka:198-guerin-pra-stirap21staticcomp,ka:198-bohmer-pra-stirap21exp,kr:215-bergmannetal-jchemphys-revstirap}, where the available phase control is limited. In addition phase modulated 2+1 STIRAP is
naturally resilient to non-Markovian noise inducing slow fluctuations~\cite{note:suppl-mat} of the energy splittings. This corresponds to fluctuating detunings, correlated as   
$\tilde{\delta}_p = 2 \,\tilde{\delta}$ in the transmon of Ref.~\cite{ka:212-rigettisteffen-prb-trasmonshapphire}. The part of this line
contained in the high efficiency region of the $(\tilde{\delta},\tilde{\delta}_p)$ plane corresponds 
to $T_2^\prime \sim 1\,\mu\mathrm{s}$, which sets a figure for the resilience of the protocol to 
non Markovian dephasing. Quite interestingly a suitably asymmetric drive with ratio $r:=\text{Max}[\Omega_p(t)]/\text{Max}[\Omega_s(t)] > 1$ enlarges
the stability region in a way that low-frequency correlated noise affecting the device  
is dynamically decoupled [see also Fig.\ref{fig:diamond}(c)].}  

%\section{Conclusioni}
In summary we have shown how to design reliable multilevel control in $\Lambda$ configuration by 
2+1 STIRAP. The key ingredient is a new control scheme which uses pulses with suitable slowly-varying modulated phases, 
Eqs.(\ref{eq:chirps},\ref{eq:new-chirps}).  
We obtained a unique strategy allowing to operate with last generation AAs, where symmetries 
enforce selection rules preventing a resonant pump field to be coupled directly. It can be easily implemented in such devices  with available microwave electronics~\cite{ka:214-pechalwallraff-prx-singlephoton}, yielding
$\sim 100\%$ efficiency. It is worth stressing that phase control is necessary to guarantee the important property of robustness to the same level of conventional STIRAP.

We finally mention that STIRAP has been very recently observed in the so called Ladder
configuration~\cite{ka:216-kumarparaoanu-natcomm-stirap,ka:216-xuzhao-ncomm-stirap}, which 
is more easily implemented in last-generation AAs.
It involves a two-photon absorption process, whereas $\Lambda$-STIRAP implements a 
coherent absorption-emission cycle. This latter is a fundamental building block for advanced 
control in highly integrated architectures, thereby it would have an impact on applications. 
Phase control exalts in a natural way the advantages of last-generation superconducting AAs, 
where it opens new perspectives for advanced quantum control. Our work may be extended in these directions using optimal control theory tools.

%We mention that the 2+1 $\Lambda$ scheme without phase control has been used to detect coherent population
%trapping in superconductors~\citep{ka:210-kellypappas-prl-cpt}, with efficiency $\gtrsim 60 \%$ only.
\begin{comment}
Our phase control scheme depends on rather general spectral properties of the device. In particular  
we have proposed two solutions, suitable respectively in highly anharmonic flux qudits  
and in the nearly harmonic transmon devices, both displaying large decoherence times. 

\usepackage[normalem]{ulem}
\sout{of performing demanding protocols as quantum processing, information transfer and entanglement generation~\cite{ka:208-weinori-prl-stirapqcomp,ka:204-yang-prl-fluxstirap,ka:204-kis-prb-fluxstirap} between remote nodes,  
novel tasks involving several $\Lambda$-STIRAP-like control cycles~\cite{ka:208-dmoller-prl-adiabgates},
and to develop control protocols~\cite{kr:207-kral-rmp-controladpass,kr:215-bergmannetal-jchemphys-revstirap} for highly integrated architectures~\cite{ka:214-mohebbicory-japph-arraymicrostrip,kka:216-brechtschoelkopf-natqinfo}, 
handling special design constraints~\cite{ka:215-distefano}. }

\end{comment}
We thank A. D'Arrigo, K. Bergmann, D. Esteve, R. Fazio, S. Guerin, G.S. Paraoanu, Yu. Pashkin, M. Paternostro and J.S. Tsai for useful discussions. 

%We acknowledge partial support by Centro Siciliano di Fisica Nucleare e Struttura della Materia and by MIUR through Grant. No.PON02 00355 3391233, "Tecnologie per l'ENERGia e l'Efficienza energETICa- ENERGETIC".
%\bibliography{stirap,misc,atom-cavity,superconducting-qubits,misc}

\pagebreak
\widetext
\begin{center}
\textbf{\large Supplementary Material}
\end{center}
%%%%%%%%%% Merge with supplemental materials %%%%%%%%%%
%%%%%%%%%% Prefix a "S" to all equations, figures, tables and reset the counter %%%%%%%%%%
\setcounter{equation}{0}
\setcounter{figure}{0}
\setcounter{table}{0}
\setcounter{page}{1}
\makeatletter
\renewcommand{\theequation}{S\arabic{equation}}
\renewcommand{\thefigure}{S\arabic{figure}}
\renewcommand{\bibnumfmt}[1]{[S#1]}
\renewcommand{\citenumfont}[1]{S#1}

\section{Effective Hamiltonian by the Magnus expansion}
We consider a time-dependent Hamiltonian $H(t)$. Our goal is to find an effective Hamiltonian
$\tilde{H}(t)$ capturing the dynamics on a coarse grained scale, defined by the small 
but finite time interval $\Delta t$. To this end we write 
\begin{equation}
\label{eq:time-evolution}
U(t+\Delta t;t) = {\cal T} \text{e}^{-i \int_t^{t+\Delta t} dt^\prime H(t^\prime)}
= \prod_{k=1}^n \text{e}^{-i H_k \,\delta t_k}
\end{equation}
where in the latter product we consider 
$n \to \infty$ time slices  $\delta t_k$ with 
$\sum_{k=1}^n \delta t_k = \Delta t$, we define 
$H_k := H(t_k)$ with $t_k$ belonging to the $k$-th time slice, and  
keep ordered in time the exponential operators. 
By repeated application of the Campbell-Baker-Hausdorff relation
\begin{equation}
\text{e}^A \text{e}^B = \text{exp}\{A+B+\frac{1}{2}[A,B]+...\}
\end{equation}
we can write $U(t+\Delta t;t) = \mathrm{e}^{- i  \tilde{H}_{\Delta t}(t) \Delta t}$, where
$\tilde{H}_{\Delta t}(t)$ is given up to second order by 
\begin{equation}
\label{eq:AE}
\tilde{H}_{\Delta t}(t) = \frac{1}{\Delta t} \int_{t}^{t+\Delta t} dt^\prime H(t^\prime) -\frac{i}{2 \Delta t} \int_{t}^{t+\Delta t} dt^\prime\int_{t}^{t^\prime} 
dt^{\prime\prime} [H(t^\prime),H(t^{\prime\prime})]
\end{equation}
We shall see that in our case this expression can be approximated by a $\Delta t$ independent one, i.e. $\tilde{H}_{\Delta t}(t) \simeq \tilde{H}(t)$.  The resulting averaged 
$U(t+\Delta t;t) = \mathrm{e}^{- i  \tilde{H}(t) \Delta t}$ allows to approximate $U(t,t_0) \approx {\cal T} \text{e}^{-i \int_{t_0}^t dt^\prime \tilde{H}(t^\prime)}$. In this way, $\tilde{H}(t)$ is identified as  an effective Hamiltonian capturing the dynamics in a coarse grained fashion.

We carried out our calculations in the interaction picture of Eqs. (4,7) of the main text. In our case each component of the pump pulse is far detuned from each transition, allowing us to take $\Delta t \delta_{ij}^k \gg 1$, where $\delta_{ij}^k:=|\omega_{ij}-\omega_{pk}|$. Since 
%all the other energy scales of the problem are orders of magnitudes smaller than the detunings, i.e. 
$\Omega_r , 
|\dot{\phi}_{pk'}(t)|, 1/T \ll \delta_{ij}^k$, we are allowed to choose $\Delta t$ such that
%\begin{equation}
$\Delta t/T, \Delta t \Omega_r, \Delta t |\dot{{\phi}}_{p2,s}(t)| \ll 1$.
%\end{equation}
Then the effective Hamiltonians Eqs. (5,8) of the main text are obtained from Eq. \ref{eq:AE} by bringing out of the integrals all the slowly varying terms and subsequently neglecting terms of order $(\delta_{ij}^k \Delta t)^{-1}$ or higher, so that $\Delta t$ will not appear in $\tilde{H}$. The physical quantities that this procedure yields are Stark shifts in the diagonal elements and amplitudes for two-photon processes in the off-diagonal elements of $\tilde{H}$.

\section{Superconducting circuits}
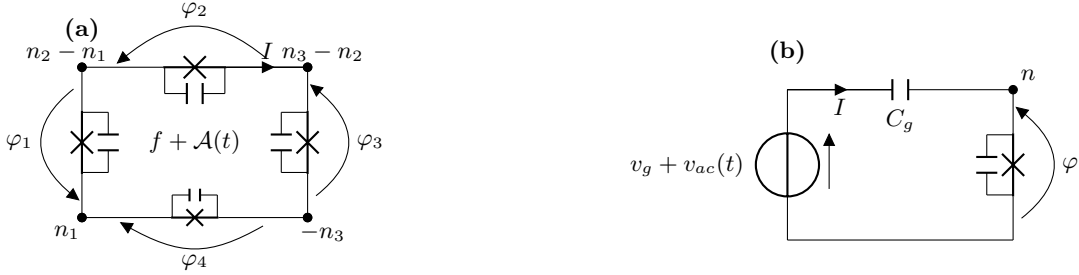
\begin{figure}[t!]
\begin{minipage}{0.48 \columnwidth}
\begin{circuitikz}
  \draw (2.5,0) to[barrier,v_>=$\varphi_3$,*-*] (2.5,2);
  \draw (2.15,0.6) to[C] (2.15,1.4);
  \draw (2.15,0.6) -- (2.5,0.6);
  \draw (2.15,1.4) -- (2.5,1.4);
  \draw (2.5,2) to[barrier,v_>=$\varphi_2$,i<_=$I$,*-*] (-0.5,2);
  \draw (0.6,1.65) to[C] (1.4,1.65);
  \draw (0.6,1.65) -- (0.6,2);
  \draw (1.4,1.65) -- (1.4,2);
  \draw (-0.5,0) to[barrier,v^=$\varphi_1$,*-*] (-0.5,2);
  \draw (-0.15,0.6) to[C] (-0.15,1.4);
  \draw (-0.15,0.6) -- (-0.5,0.6);
  \draw (-0.15,1.4) -- (-0.5,1.4);
  \ctikzset{bipoles/barrier/height=.45}
\ctikzset{bipoles/barrier/width=.45} 
\ctikzset{bipoles/capacitor/height=.12} 
\ctikzset{bipoles/capacitor/width=.10} 
  \draw (-0.5,0) to[barrier,v=$\varphi_4$,*-*] (2.5,0);
  \draw (0.7,0.3) to[C] (1.3,0.3);
  \draw (0.7,0.3) -- (0.7,0);
  \draw (1.3,0.3) -- (1.3,0);
  \draw (1,1) node {$f+{\cal A}(t)$};
  \draw (-0.7,2.2) node {$n_2-n_1$};
  \draw (2.7,2.2) node {$n_3-n_2$};
  \draw (-0.7,-0.2) node {$n_1$};
  \draw (2.7,-0.2) node {$-n_3$};
  \draw (-0.5,2.5) node {$\textbf{(a)}$};
\end{circuitikz}
\end{minipage}
\begin{minipage}{0.48 \columnwidth}
\begin{circuitikz}
  \draw (2.5,0) to[barrier,v_>=$\varphi$,-*] (2.5,2);
  \draw (2.15,0.6) to[C] (2.15,1.4);
  \draw (2.15,0.6) -- (2.5,0.6);
  \draw (2.15,1.4) -- (2.5,1.4);
  \draw (2.5,2) to[C=$C_g$,i<=$I$] (-0.5,2);
  \draw (-0.5,2) to[european voltage source,v<=$ $,l_=$v_g+v_{ac}(t)$] (-0.5,0);
  \draw (-0.5,0) -- (2.5,0);
  \draw (-0.5,2.5) node {$\textbf{(b)}$};
  \draw (2.7,2.2) node {$n$};
\end{circuitikz}
\end{minipage}
\caption{
\label{fig:circuits}
Schematics for the superconducting devices considered in the simulations, where $\varphi_i$ 
are superconducting phases across the junctions and the canonically conjugated 
$n_i$ are the numbers of extra Cooper pairs in the islands. (a) The flux-based device made out a four Josephson junctions SQUID loop. The three bigger junctions have Josephson energy $E_J$ and parallel capacitance $C$ while the smaller one has Josephson energy and capacitance smaller by a factor $\alpha$. (b) The Cooper pair box is a superconducting island 
separated from the circuit by a single Josephson junction of energy $E_J$ and capacitance $C$. The transmon is a capacitively-shunted Cooper pair box with  $E_J/E_C \gg 1$
strongly coupled to an electromagnetic transmission line resonator, which allows quantum
non-demolition measurements.}
\end{figure}

The superconducting circuits considered in the text are depicted in Fig. \ref{fig:circuits} (a) (flux-based device) and (b) (transmon). For both we will give the undriven Hamiltonian 
$H_0$ and the coupling operator ${\cal Q}$ entering the control Hamiltonian $H_C := {\cal Q} {\cal A}(t)$ [See Eq. (2) in main text] with the associated selection rules and symmetries.
\subsection{Flux Qubit}
The flux qubit is made out of a SQUID superconducting loop with four junction. Three of them have equal Josephson energies $E_J$ and capacitances $C$, whereas the fourth one is smaller by a factor $\alpha$. The Hamiltonian exressed in terms of the three independent phases 
$\boldsymbol{\varphi}:=(\varphi_1,\varphi_2,\varphi_3)$ and the associated charges reads
\begin{equation}
H = -E_J \sum_j^3 \cos \hat{\varphi}_j - \alpha E_J \cos \big\{ \sum_j^3 \hat{\varphi}_j - 2\pi [f + {\cal A}(t)] \big\} + 4 \frac{E_C}{1+3\alpha} \big[ (1+2\alpha) \sum_j^3 \hat{n}_i^2 -2\alpha \sum_{i\neq j}^3 \hat{n}_i \hat{n}_j \big]
\end{equation}
where we neglected the parasitic capacitances~\cite{ka:214:sternvion-prl-flux}. 
Here $E_C=e^2/2C$ is the single electron charging energy and $f = \Phi_b/\Phi_0$ is the external
magnetic flux bias $\Phi_b$ in units of the flux quantum $\Phi_0=h/(2e)$, and  
we work at the symmetry point $f=1/2$.
Physically the control is a small magnetic flux $\Phi_{ac}(t)$ added to $\Phi_b$, and 
${\cal A}(t) = \Phi_{ac}(t)/\Phi_0$. 
By expanding to first order in ${\cal A}$ we find the 
coupling operator ${\cal Q} = 2\pi \alpha E_J \sin \big( \sum_j^3 \hat{\varphi}_j) = 2\pi E_J \hat{I}/I_c$, where $\hat{I}$ is the loop current operator and $I_c$ the critical current of the big junctions. 
For $f=1/2$ the bare Hamiltonian $H_0$ enjoys a symmetry with respect to the parity operator ${\cal P}_{\varphi}\ket{\boldsymbol{\varphi}}=\ket{-\boldsymbol{\varphi}}$. As a consequence, eigenfunctions 
$\ket{n}$ of $H_0$ can be chosen with a definite symmetry, i.e. $\psi_n(-\boldsymbol{\varphi}) = (-1)^n \psi_n(\boldsymbol{\varphi})$, where $n$ labels eigenenergies in increasing order, implying the selection rule for the odd parity coupling operator ${\cal Q}$
\begin{equation}
\label{eq:selection-rule}
\bra{n}{\cal Q}\ket{m} \propto 1 - (-1)^{m+n}
\end{equation}
In our simulations, we employed values $E_C/2\pi=4 \text{GHz}$, $E_J = 50 E_C$ and $\alpha=0.54$ as in Ref~\onlinecite{ka:211-bylander-natphys}, and used $13^3$ charge states to diagonalize $H_0$. 

\ctikzset{bipoles/capacitor/height=.2} 
\ctikzset{bipoles/capacitor/width=.15} 
\subsection{Transmon}
The transmon can be modelled by a Cooper pair box [see Fig. \ref{fig:circuits} (b)] in the $E_J/E_C \gg 1$ regime~\cite{ka:207-koch-pra-transmon}. Here $E_J$ is the Josephson energy of the junction of the box, while the 
charging energy $E_C = e^2/2C_{tot}$ involves the total capacitance $C_{tot}=C+C_g$. The Hamiltonian reads
\begin{equation}
\label{eq:transmon}
H = -E_J \cos \hat{\varphi} + 4 E_C (\hat{n}-q_g-{\cal A}(t))^2
\end{equation}
where $q_g = C_{tot} v_g/2e$ is a bias parameter. The control field ${\cal A}(t) = C_{tot} v_{ac}(t)/2e$ couples to operator ${\cal Q} = - 8 E_C (\hat{n}-q_g)$. 
Biasing the device at $q_g=1/2$ the Hamiltonian $H_0$ is symmetric with respect to the 
charge-parity operator ${\cal P}_n = \sum_m \ketbra{1-m}{m}$, implying that also in this case 
a selection rule of the kind Eq.(\ref{eq:selection-rule}). In the simulations 
we employed values $E_C/2\pi=0.212 \text{GHz}$, $E_J/E_C=49$ as in Ref.~\cite{ka:212-rigettisteffen-prb-trasmonshapphire}, 
using $100$ charge states to diagonalize $H_0$.

\section{Effects of noise}

Different noise sources act on the devices, but independently on their nature they produces essentially two distinct classes of effects~\cite{ka:205-falci-prl,kr:214-paladino-rmp}. 
Environmental modes with frequencies comparable to Bohr or Rabi frequencies act as sources of Markovian quantum noise, whose leading low-temperature effects in STIRAP are spontaneous decay, the associated secular dephasing and field-induced absorption~\cite{ka:213-falci-prb-stirapcpb,ka:195-geva-jorchemphys-gmerabi}. At lower frequencies noise in the solid state is non-Markovian and exhibits a $1/f^\alpha$ behavior. The leading effect is a pure dephasing, analogous 
to inhomogeneous broadening produced by a classical noise source, and it is effectively described by a stochastic external drive with the desired spectral features~\cite{ka:205-falci-prl,kr:214-paladino-rmp}. 

According to this picture, we describe Markovian quantum noise by a 
standard dissipator ${\mathcal L}_D$ term  in a Lindblad Master Equation
\begin{equation}
\label{eq:lowfreq}
\dot{\rho}_f(t|\mathbf{\tilde{x}})
=-i[H(\mathbf{\tilde{x}}(t)),\rho_f(t|\mathbf{\tilde{x}})] + {\mathcal L}_D \rho_f(t|\mathbf{\tilde{x}})
\end{equation}
where we also account for the effect of non-Markovian noise by allowing the parametric dependence 
on the classical stochastic processes $\{\mathbf{\tilde{x}}(t)\}$. 
%Under reasonable conditions these latter can be tought to affect only the system Hamiltonian $H$. The full noisy dynamics is obtained by averaging over the classical stochastic processes. 
Averaging over this latter yields the full noisy dynamics
\begin{equation}
\label{eq:average}
\rho(t) = \int {\cal D} \mathbf{\tilde{x}}(t) \,P[\mathbf{\tilde{x}}(t)]\, \rho(t|\mathbf{\tilde{x}})
\end{equation}
%An important simplification comes from the fact that low-frequency noise has a $\sim 1/f^\alpha$ behavior, thereby the relevant part of its power spectrum contains frequencies $\ll 1/T$. As a consequence its effects can be seen as quasistatic fluctuations of the energy levels $\Delta E_i$ and of the dipole matrix elements $\Delta {\cal Q}_{ij}$. 
If low-frequency noise has a $\sim 1/f^\alpha$ spectrum the main effect is to induce 
quasistatic stray bias $\mathbf{\tilde{x}}$ of the artificial atom. The path integral Eq.(\ref{eq:average}) 
can therefore be evaluated in the Static Path Approximation(SPA)~\cite{ka:205-falci-prl,kr:214-paladino-rmp}, reducing to an ordinary integration over 
random variables $\mathbf{\tilde{x}}$. These latter are moreover Gaussian distributed if we assume that they are due 
to many uncorrelated microscopic sources. 

We apply this recipe to 2+1 STIRAP in HNP devices. The emerging important qualitative issue
is that noisy three-level dynamics is fully characterized by decoherence in the "trapped" 
(or qubit) subspace only, plus information on the Hamiltonian of the device alone, a   
results also obtained conventional STIRAP~\cite{ka:213-falci-prb-stirapcpb}. 
Indeed stray bias due to low-frequency noise determine
fluctuations of energies $\Delta E_i$ and of matrix elements $\Delta {\cal Q}_{ij}$. They 
translate respectively in fluctuations of the detunings $\delta=\Delta E_1-\Delta E_0$ and $\delta_p =\Delta E_2-\Delta E_0$ and in fluctuations of the Rabi frequency $\tilde{\Omega}_0$.
The sensitivity to such parameters has been extensively studied~\cite{kr:201-vitanov-advatmolopt}: 
while fluctuations  $\tilde{\Omega}_0$ are irrelevant for STIRAP, fluctuations of detunings are important.  
Therefore the relevant open system dynamics turns out to be described by a Lindblad Master Equation with the structure $\dot{\rho}=i[\rho,H(\delta,\delta_p)] + {\mathcal L}_D \rho$ where $H$ depends on fluctuations $(\delta,\delta_p)$ induced by stray bias $\mathbf{\tilde{x}}$. 
In Fig.3(c) of the main text we plot the efficiency of the protocol vs stray $(\delta,\delta_p)$: efficiency is large if fluctuations do not let the system  
diffuse out of the central diamond region. In particular the protocol is critically sensitive to fluctuations 
of the two-photon detuning $\delta$, i.e. of the "qubit" splitting $E_1 - E_0$. 

Concerning quantum noise, ${\mathcal L}_D \rho$  includes in principle the various decay rates and associated excitation and secular dephasing processes, but in practice again the "qubit" spontaneous decay only has to be accounted for, i.e. the only relevant term 
turns out to be  
%\begin{equation}\label{eq:ME}
${\mathcal L}_D \rho =
- \frac{1}{2T_1}([\sigma_+ \sigma_-,\rho]-2\sigma_-\rho\sigma_+)$
%\end{equation}
where $1/T_1$ is the spontaneous decay rate of the qubit, and the Lindblad operators are the 
corresponding lowering and raising operators $\sigma_-=\sigma_+^{\dagger}=\ketbra{0}{1}$. Indeed selection rules suppress  
$2 \leftrightarrow 0$ processes, whereas we can estimate the transitions rate 
$\Gamma_{2 \to 1} = k/T_1$, where $k = O(1)$ depends on features of the device, and 
directly check that in the devices of interest it has no impact on the results. This is due to the fact that $\ket{2}$ is depopulated when STIRAP is successful. The same holds true for field induced absorption, since $\Omega_s$ ($\Omega_{pk}$) act when $\ket{1}$ ($\ket{0}$) are depopulated. Notice finally that we did not include in the dissipator extra pure dephasing  terms since they are accounted for by the average over non-Markovian quasistatic fluctuations.
\begin{figure}[t]
\includegraphics[width=0.3\columnwidth]{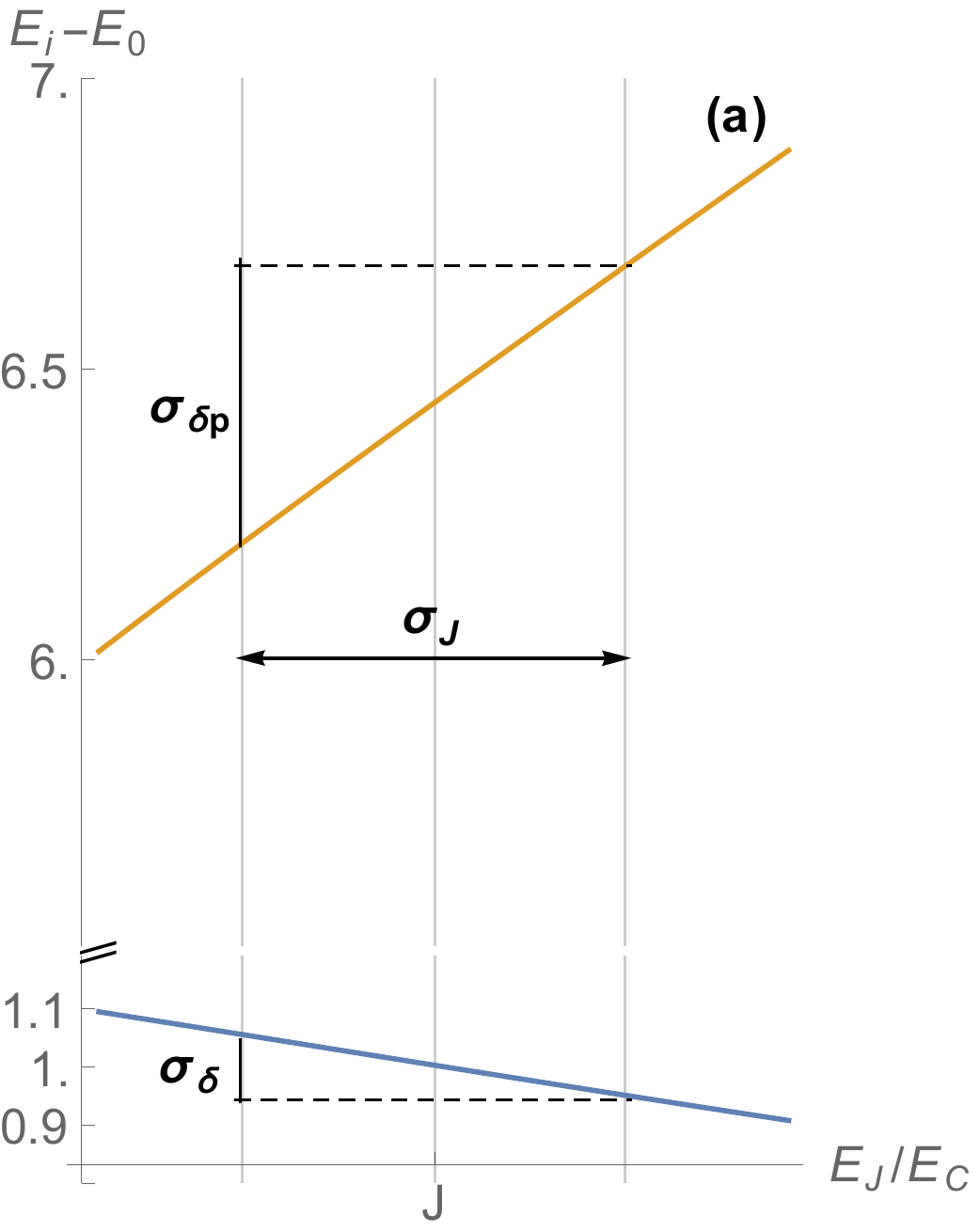}
\includegraphics[width=0.3\columnwidth]{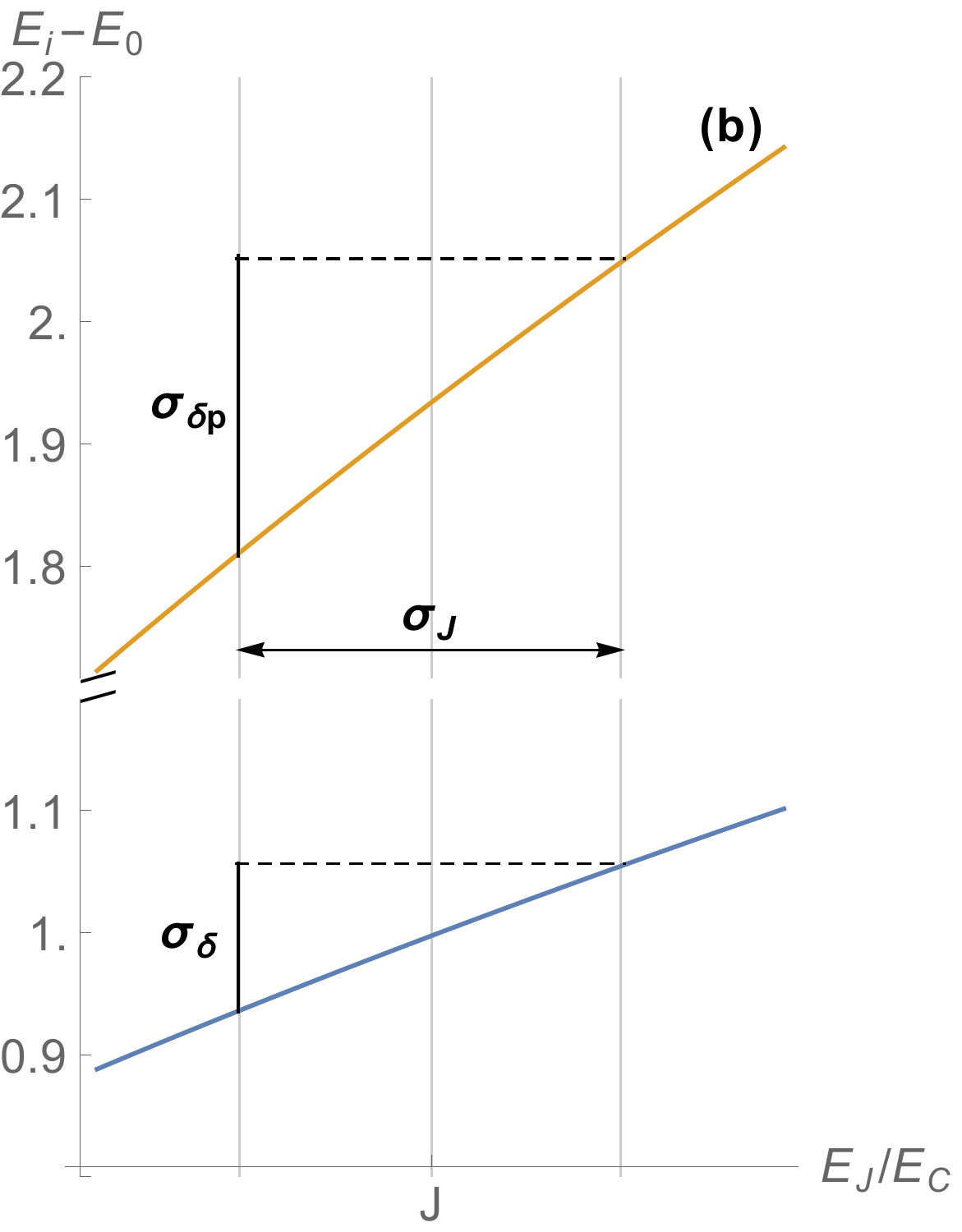}
\caption{
Bohr frequencies $E_1 - E_0$ (lower coloured curve) and $E_2 - E_0$ vs $E_J/E_C$ (upper coloured curve) around a bias point $J$ for a flux qudit (a) and a transmon (b). It is seen that Bohr frequencies vary linearly with respect to the fluctuating parameter $E_J/E_C$. Variances $\sigma_{\delta}$ and $\sigma_{\delta_p}$ are indicated in correspondence of a noise variance $\sigma_J$. Correlation of detuning fluctuations, i.e. 
$\delta_p = a \delta$, is determined through $a= [\partial (E_2-E_0)/\partial (E_J/E_C)]/
[\partial (E_1-E_0)/\partial (E_J/E_C)]$, where the derivative is evaluated at the bias point $J$.
\label{fig:bandstructure}
}
\end{figure}

We now briefly discuss how figures of noise can be extracted from experiments on qubits, referring to the flux qudit of Ref. \onlinecite{ka:211-bylander-natphys}. 
The dominant source of decoherence is flux noise. The low-frequency part of its spectrum 
$S_\Phi(\omega)$, though, has minimal effect since at the symmetry point $f=1/2$, energy fluctuations are {\em quadratic} in the small corresponding stray bias $\tilde{x}_1(t)=\tilde{f}(t)$.
Instead its high-frequency components determine the qubit $T_1=12 \,\mu\mathrm{s}$
and the rate $\Gamma_{2 \to 1}$. For this latter we estimated 
$k=  [Q_{21}/Q_{10}]^2\, \,[S_\Phi[(E_2-E_1)/\hbar]/S_\Phi[(E_1-E_0)/\hbar]
\lesssim 1$, taking for $S_\Phi(\omega)$ the linear behavior observed for quantum noise~\cite{ka:211-bylander-natphys}, and verified that it  does not affect the dynamics.
Subdominant noise sources for flux qudits are critical current and charge noise.
They do not produce relaxation at the symmetry point, but they are the main source of non-Markovian pure dephasing. The induced energy fluctuations are {\em linear}
in the corresponding stray bias $\mathbf{\tilde{x}}(t)$ (see Fig.\ref{fig:bandstructure}),
determining Gaussian suppression of the qubit coherence~\cite{kr:214-paladino-rmp}. 
The power spectrum $S_{\mathbf x}(\omega)\sim 1/\omega$ has been measured from the qubit dynamics, over several frequency decades~\cite{ka:211-bylander-natphys}. 
Therefore we can safely use the SPA result for the qubit coherences decay $\propto \mathrm{e}^{-\sigma_{\delta}^2 t^2/2}$ and 
extract from the measured non-Markovian qubit pure dephasing rate $1/T_2^\prime:=1/T_2^*-1/(2T_1)$ the variance 
$\sigma_\delta=\sqrt{2}/T_2^\prime$ of the 
two-photon detuning in STIRAP. 
Fluctuations of $\delta_p$, i.e. of $E_2-E_0$, are easily found from the parametric dependence on the external bias of the calculated spectrum of the device 
(see Fig.\ref{fig:bandstructure}). Notice that since each source of noise induces a single stray bias $\tilde{x}_i$, fluctuations of detunings are correlated. For HNP devices subdominant noise sources induce fluctuations $\delta_p = a \delta$. The constant $a$ depends on the band structure of the device
(see Fig.~\ref{fig:bandstructure}) and in particular $a=-4.5$ in the flux qudit of Ref.~\onlinecite{ka:211-bylander-natphys} refers to critical current and charge noise. 
 
Similar considerations hold for the transmon of Ref.~\cite{ka:212-rigettisteffen-prb-trasmonshapphire} $a \simeq 2$ and $k\simeq 2$, symmetry suppresses low-frequency charge noise, and subdominant noise as flux and critical current noise lead to pure dephasing $1/T^\prime_2$ which is very small. 

Numerics have been carried out through a Montecarlo quantum jump approach accounting for Markovian noise, by averaging over $10^4$ trajectories. Non-Markovian noise has been taken into account by a further average: we impose $\delta_p = a  \delta$ and sample $\delta$ from its Gaussian distribution for each trajectory. Variances we used are $\sigma_{\delta} = 4.1 \times 10^{-3}\,\Omega_0= -0.22 \,\sigma_{\delta_p}$ for flux qubit ($\Omega_0 = 20 \,\text{MHz}$) and $\sigma_{\delta} = 2 \times 10^{-4} \,\Omega_0 = 0.5 \,\sigma_{\delta_p}$ in the transmon.
By inspection of Fig.3(c) of the main text it is clear that STIRAP in HNP devices is robust against such low.frequency fluctuations for both devices, as it also results from  simulations.
%\bibliography{../stirap,misc,atom-cavity,../superconducting-qubits}

\begin{thebibliography}{50}%
\makeatletter
\providecommand \@ifxundefined [1]{%
 \@ifx{#1\undefined}
}%
\providecommand \@ifnum [1]{%
 \ifnum #1\expandafter \@firstoftwo
 \else \expandafter \@secondoftwo
 \fi
}%
\providecommand \@ifx [1]{%
 \ifx #1\expandafter \@firstoftwo
 \else \expandafter \@secondoftwo
 \fi
}%
\providecommand \natexlab [1]{#1}%
\providecommand \enquote  [1]{``#1''}%
\providecommand \bibnamefont  [1]{#1}%
\providecommand \bibfnamefont [1]{#1}%
\providecommand \citenamefont [1]{#1}%
\providecommand \href@noop [0]{\@secondoftwo}%
\providecommand \href [0]{\begingroup \@sanitize@url \@href}%
\providecommand \@href[1]{\@@startlink{#1}\@@href}%
\providecommand \@@href[1]{\endgroup#1\@@endlink}%
\providecommand \@sanitize@url [0]{\catcode `\\12\catcode `\$12\catcode
  `\&12\catcode `\#12\catcode `\^12\catcode `\_12\catcode `\%12\relax}%
\providecommand \@@startlink[1]{}%
\providecommand \@@endlink[0]{}%
\providecommand \url  [0]{\begingroup\@sanitize@url \@url }%
\providecommand \@url [1]{\endgroup\@href {#1}{\urlprefix }}%
\providecommand \urlprefix  [0]{URL }%
\providecommand \Eprint [0]{\href }%
\providecommand \doibase [0]{http://dx.doi.org/}%
\providecommand \selectlanguage [0]{\@gobble}%
\providecommand \bibinfo  [0]{\@secondoftwo}%
\providecommand \bibfield  [0]{\@secondoftwo}%
\providecommand \translation [1]{[#1]}%
\providecommand \BibitemOpen [0]{}%
\providecommand \bibitemStop [0]{}%
\providecommand \bibitemNoStop [0]{.\EOS\space}%
\providecommand \EOS [0]{\spacefactor3000\relax}%
\providecommand \BibitemShut  [1]{\csname bibitem#1\endcsname}%
\let\auto@bib@innerbib\@empty
%</preamble>
\bibitem [{\citenamefont {Nielsen}\ and\ \citenamefont
  {Chuang}(2010)}]{kb:210-nielsenchuang}%
  \BibitemOpen
  \bibfield  {author} {\bibinfo {author} {\bibfnamefont {M.}~\bibnamefont
  {Nielsen}}\ and\ \bibinfo {author} {\bibfnamefont {I.}~\bibnamefont
  {Chuang}},\ }\href@noop {} {\emph {\bibinfo {title} {Quantum Computation and
  Quantum Information}}}\ (\bibinfo  {publisher} {Cambridge Univ. Press,
  Cambridge},\ \bibinfo {year} {2010})\BibitemShut {NoStop}%
\bibitem [{\citenamefont {Timoney}\ \emph {et~al.}(2011)\citenamefont
  {Timoney}, \citenamefont {Baumgart}, \citenamefont {Johanning}, \citenamefont
  {Varon}, \citenamefont {Plenio}, \citenamefont {Retzker},\ and\ \citenamefont
  {Wunderlich}}]{ka:211-timoney-nature-dressed}%
  \BibitemOpen
  \bibfield  {author} {\bibinfo {author} {\bibfnamefont {N.}~\bibnamefont
  {Timoney}}, \bibinfo {author} {\bibfnamefont {I.}~\bibnamefont {Baumgart}},
  \bibinfo {author} {\bibfnamefont {M.}~\bibnamefont {Johanning}}, \bibinfo
  {author} {\bibfnamefont {A.~F.}\ \bibnamefont {Varon}}, \bibinfo {author}
  {\bibfnamefont {M.~B.}\ \bibnamefont {Plenio}}, \bibinfo {author}
  {\bibfnamefont {A.}~\bibnamefont {Retzker}}, \ and\ \bibinfo {author}
  {\bibfnamefont {C.}~\bibnamefont {Wunderlich}},\ }\href {\doibase
  {10.1038/nature10319}} {\bibfield  {journal} {\bibinfo  {journal} {Nature}\
  }\textbf {\bibinfo {volume} {476}},\ \bibinfo {pages} {185} (\bibinfo {year}
  {2011})}\BibitemShut {NoStop}%
\bibitem [{\citenamefont {You}\ and\ \citenamefont
  {Nori}(2011)}]{kr:211-younori-nature-multilevel}%
  \BibitemOpen
  \bibfield  {author} {\bibinfo {author} {\bibfnamefont {J.~Q.}\ \bibnamefont
  {You}}\ and\ \bibinfo {author} {\bibfnamefont {F.}~\bibnamefont {Nori}},\
  }\href {\doibase 10.1038/nature10122} {\bibfield  {journal} {\bibinfo
  {journal} {Nature}\ }\textbf {\bibinfo {volume} {{\bf 474}}},\ \bibinfo
  {pages} {589} (\bibinfo {year} {2011})}\BibitemShut {NoStop}%
\bibitem [{\citenamefont {Lanyon}\ \emph {et~al.}(2009)\citenamefont {Lanyon},
  \citenamefont {Barbieri}, \citenamefont {Almeida}, \citenamefont {Jennewein},
  \citenamefont {Ralph}, \citenamefont {Resch}, \citenamefont {Pryde},
  \citenamefont {O’Brien},\ and\ \citenamefont
  {White}}]{ka:209-lanyonwhite-natphys-quditprocessing}%
  \BibitemOpen
  \bibfield  {author} {\bibinfo {author} {\bibfnamefont {B.~P.}\ \bibnamefont
  {Lanyon}}, \bibinfo {author} {\bibfnamefont {M.}~\bibnamefont {Barbieri}},
  \bibinfo {author} {\bibfnamefont {M.~P.}\ \bibnamefont {Almeida}}, \bibinfo
  {author} {\bibfnamefont {T.}~\bibnamefont {Jennewein}}, \bibinfo {author}
  {\bibfnamefont {T.~C.}\ \bibnamefont {Ralph}}, \bibinfo {author}
  {\bibfnamefont {K.~J.}\ \bibnamefont {Resch}}, \bibinfo {author}
  {\bibfnamefont {G.~J.}\ \bibnamefont {Pryde}}, \bibinfo {author}
  {\bibfnamefont {A.}~\bibnamefont {O’Brien}, \bibfnamefont {Jeremy
  L.~Gilchrist}}, \ and\ \bibinfo {author} {\bibfnamefont {A.~G.}\ \bibnamefont
  {White}},\ }\href@noop {} {\bibfield  {journal} {\bibinfo  {journal} {Nat. Phys.}\ }\textbf {\bibinfo {volume} {5}} (\bibinfo {year}
  {2009})}\BibitemShut {NoStop}%
\bibitem [{\citenamefont {Pachos}(2012)}]{kb:209-pachos-topqcomp}%
  \BibitemOpen
  \bibfield  {author} {\bibinfo {author} {\bibfnamefont {J.}~\bibnamefont
  {Pachos}},\ }\href@noop {} {\emph {\bibinfo {title} {Introduction to
  Topological Quantum Computation}}}\ (\bibinfo  {publisher} {Cambridge
  University Press},\ \bibinfo {year} {2012})\BibitemShut {NoStop}%
\bibitem [{\citenamefont {Kimble}(2008)}]{ka:208-kimble-nature-qinternet}%
  \BibitemOpen
  \bibfield  {author} {\bibinfo {author} {\bibfnamefont {H.}~\bibnamefont
  {Kimble}},\ }\href@noop {} {\bibfield  {journal} {\bibinfo  {journal}
  {Nature}\ }\textbf {\bibinfo {volume} {453}},\ \bibinfo {pages} {1023}
  (\bibinfo {year} {2008})}\BibitemShut {NoStop}%
\bibitem [{\citenamefont {D'Arrigo}\ \emph {et~al.}(2014)\citenamefont
  {D'Arrigo}, \citenamefont {Lo~Franco}, \citenamefont {Benenti}, \citenamefont
  {Paladino},\ and\ \citenamefont {Falci}}]{ka:214-darrigo-annals-hiddenent}%
  \BibitemOpen
  \bibfield  {author} {\bibinfo {author} {\bibfnamefont {A.}~\bibnamefont
  {D'Arrigo}}, \bibinfo {author} {\bibfnamefont {R.}~\bibnamefont {Lo~Franco}},
  \bibinfo {author} {\bibfnamefont {G.}~\bibnamefont {Benenti}}, \bibinfo
  {author} {\bibfnamefont {E.}~\bibnamefont {Paladino}}, \ and\ \bibinfo
  {author} {\bibfnamefont {G.}~\bibnamefont {Falci}},\ }\href {\doibase
  10.1016/j.aop.2014.07.021} {\bibfield  {journal} {\bibinfo  {journal} {Ann. Phys.}\ }\textbf {\bibinfo {volume} {350}},\ \bibinfo {pages} {211}
  (\bibinfo {year} {2014})}\BibitemShut {NoStop}%
\bibitem [{\citenamefont {Orieux}\ \emph {et~al.}(2015)\citenamefont {Orieux},
  \citenamefont {D'Arrigo}, \citenamefont {Ferranti}, \citenamefont
  {Lo~Franco}, \citenamefont {Benenti}, \citenamefont {Paladino}, \citenamefont
  {Falci}, \citenamefont {Sciarrino},\ and\ \citenamefont
  {Mataloni}}]{ka:215:orieux-scirep-recoveryent}%
  \BibitemOpen
  \bibfield  {author} {\bibinfo {author} {\bibfnamefont {A.}~\bibnamefont
  {Orieux}}, \bibinfo {author} {\bibfnamefont {A.}~\bibnamefont {D'Arrigo}},
  \bibinfo {author} {\bibfnamefont {G.}~\bibnamefont {Ferranti}}, \bibinfo
  {author} {\bibfnamefont {R.}~\bibnamefont {Lo~Franco}}, \bibinfo {author}
  {\bibfnamefont {G.}~\bibnamefont {Benenti}}, \bibinfo {author} {\bibfnamefont
  {E.}~\bibnamefont {Paladino}}, \bibinfo {author} {\bibfnamefont
  {G.}~\bibnamefont {Falci}}, \bibinfo {author} {\bibfnamefont
  {F.}~\bibnamefont {Sciarrino}}, \ and\ \bibinfo {author} {\bibfnamefont
  {P.}~\bibnamefont {Mataloni}},\ }\href {\doibase {10.1038/srep08575}}
  {\bibfield  {journal} {\bibinfo  {journal} {Sci. Rep.}\ }\textbf {\bibinfo
  {volume} {5}} (\bibinfo {year} {2015})}\BibitemShut
  {NoStop}%
\bibitem [{\citenamefont {Ladd}\ \emph {et~al.}(2010)\citenamefont {Ladd},
  \citenamefont {Jelezko}, \citenamefont {Laflamme}, \citenamefont {Nakamura},
  \citenamefont {Monroe},\ and\ \citenamefont
  {O'Brien}}]{kr:210-ladd-nature-revqcomp}%
  \BibitemOpen
  \bibfield  {author} {\bibinfo {author} {\bibfnamefont {T.~D.}\ \bibnamefont
  {Ladd}}, \bibinfo {author} {\bibfnamefont {F.}~\bibnamefont {Jelezko}},
  \bibinfo {author} {\bibfnamefont {R.}~\bibnamefont {Laflamme}}, \bibinfo
  {author} {\bibfnamefont {Y.}~\bibnamefont {Nakamura}}, \bibinfo {author}
  {\bibfnamefont {C.}~\bibnamefont {Monroe}}, \ and\ \bibinfo {author}
  {\bibfnamefont {J.~L.}\ \bibnamefont {O'Brien}},\ }\href@noop {} {\bibfield
  {journal} {\bibinfo  {journal} {Nature}\ }\textbf {\bibinfo {volume} {464}},\
  \bibinfo {pages} {08812} (\bibinfo {year} {2010})}\BibitemShut {NoStop}%
\bibitem [{\citenamefont {Devoret}\ and\ \citenamefont
  {Schoelkopf}(2013)}]{kr:213-devoretschoelkopf-science}%
  \BibitemOpen
  \bibfield  {author} {\bibinfo {author} {\bibfnamefont {M.}~\bibnamefont
  {Devoret}}\ and\ \bibinfo {author} {\bibfnamefont {R.~J.}\ \bibnamefont
  {Schoelkopf}},\ }\href {\doibase 10.1126/science.1231930} {\bibfield
  {journal} {\bibinfo  {journal} {Science}\ }\textbf {\bibinfo {volume}
  {339}},\ \bibinfo {pages} {1169} (\bibinfo {year} {2013})}\BibitemShut
  {NoStop}%
\bibitem [{\citenamefont {Schoelkopf}\ and\ \citenamefont
  {Girvin}(2008)}]{kr:208-schoelkopf-nature-wiring}%
  \BibitemOpen
  \bibfield  {author} {\bibinfo {author} {\bibfnamefont {R.~J.}\ \bibnamefont
  {Schoelkopf}}\ and\ \bibinfo {author} {\bibfnamefont {S.~M.}\ \bibnamefont
  {Girvin}},\ }\href@noop {} {\bibfield  {journal} {\bibinfo  {journal}
  {Nature}\ }\textbf {\bibinfo {volume} {451}},\ \bibinfo {pages} {664}
  (\bibinfo {year} {2008})}\BibitemShut {NoStop}%
\bibitem [{\citenamefont {Macha}\ \emph {et~al.}(2014)\citenamefont {Macha},
  \citenamefont {Oelsner}, \citenamefont {Reiner}, \citenamefont {Marthaler},
  \citenamefont {Andr\`e}, \citenamefont {Sch\"on}, \citenamefont {H\"ubner},
  \citenamefont {Meyer}, \citenamefont {Il'ichev},\ and\ \citenamefont
  {Ustinov}}]{ka:214-machaustinov-ncomms-supmetamaterials}%
  \BibitemOpen
  \bibfield  {author} {\bibinfo {author} {\bibfnamefont {P.}~\bibnamefont
  {Macha}}, \bibinfo {author} {\bibfnamefont {G.}~\bibnamefont {Oelsner}},
  \bibinfo {author} {\bibfnamefont {J.-M.}\ \bibnamefont {Reiner}}, \bibinfo
  {author} {\bibfnamefont {M.}~\bibnamefont {Marthaler}}, \bibinfo {author}
  {\bibfnamefont {S.}~\bibnamefont {Andr\`e}}, \bibinfo {author} {\bibfnamefont
  {G.}~\bibnamefont {Sch\"on}}, \bibinfo {author} {\bibfnamefont
  {U.}~\bibnamefont {H\"ubner}}, \bibinfo {author} {\bibfnamefont {H.-G.}\
  \bibnamefont {Meyer}}, \bibinfo {author} {\bibfnamefont {E.}~\bibnamefont
  {Il'ichev}}, \ and\ \bibinfo {author} {\bibfnamefont {A.~V.}\ \bibnamefont
  {Ustinov}},\ }\href {\doibase 10.1038/ncomms6146} {\bibfield  {journal}
  {\bibinfo  {journal} {Nat. Commun.}\ }\textbf {\bibinfo {volume} {5}},\
  \bibinfo {pages} {5146} (\bibinfo {year} {2014})}\BibitemShut {NoStop}%
\bibitem [{\citenamefont {Mohebbi}\ \emph {et~al.}(2014)\citenamefont
  {Mohebbi}, \citenamefont {Benningshof}, \citenamefont {Taminiau},
  \citenamefont {Miao},\ and\ \citenamefont
  {Cory}}]{ka:214-mohebbicory-japph-arraymicrostrip}%
  \BibitemOpen
  \bibfield  {author} {\bibinfo {author} {\bibfnamefont {H.~R.}\ \bibnamefont
  {Mohebbi}}, \bibinfo {author} {\bibfnamefont {O.~W.~B.}\ \bibnamefont
  {Benningshof}}, \bibinfo {author} {\bibfnamefont {I.~A.~J.}\ \bibnamefont
  {Taminiau}}, \bibinfo {author} {\bibfnamefont {G.~X.}\ \bibnamefont {Miao}},
  \ and\ \bibinfo {author} {\bibfnamefont {D.~G.}\ \bibnamefont {Cory}},\
  }\href {\doibase 10.1063/1.4866691} {\bibfield  {journal} {\bibinfo
  {journal} {Jour. Appl. Phys.}\ }\textbf {\bibinfo {volume} {115}},\ \bibinfo
  {pages} {094502} (\bibinfo {year} {2014})}\BibitemShut {NoStop}%
\bibitem [{\citenamefont {Brecht}\ \emph {et~al.}(2016)\citenamefont {Brecht},
  \citenamefont {Pfaff}, \citenamefont {Wang}, \citenamefont {Chu},
  \citenamefont {Frunzio}, \citenamefont {Devoret},\ and\ \citenamefont
  {Schoelkopf}}]{ka:216-brechtschoelkopf-natqinfo}%
  \BibitemOpen
  \bibfield  {author} {\bibinfo {author} {\bibfnamefont {T.}~\bibnamefont
  {Brecht}}, \bibinfo {author} {\bibfnamefont {W.}~\bibnamefont {Pfaff}},
  \bibinfo {author} {\bibfnamefont {C.}~\bibnamefont {Wang}}, \bibinfo {author}
  {\bibfnamefont {Y.}~\bibnamefont {Chu}}, \bibinfo {author} {\bibfnamefont
  {L.}~\bibnamefont {Frunzio}}, \bibinfo {author} {\bibfnamefont {M.~H.}\
  \bibnamefont {Devoret}}, \ and\ \bibinfo {author} {\bibfnamefont {R.~J.}\
  \bibnamefont {Schoelkopf}},\ }\href {\doibase 10.1038/npjqi.2016.2}
  {\bibfield  {journal} {\bibinfo  {journal} {Npj Quantum Information}\
  }\textbf {\bibinfo {volume} {2}},\ \bibinfo {pages} {16002 EP } (\bibinfo
  {year} {2016})}\BibitemShut {NoStop}%
\bibitem [{\citenamefont {Niemczyk}\ \emph {et~al.}(2010)\citenamefont
  {Niemczyk}, \citenamefont {Deppe}, \citenamefont {Huebl}, \citenamefont
  {Menzel}, \citenamefont {Hocke}, \citenamefont {Schwarz}, \citenamefont
  {Garcia-Ripoll}, \citenamefont {Zueco}, \citenamefont {Hummer}, \citenamefont
  {Solano}, \citenamefont {Marx},\ and\ \citenamefont
  {Gross}}]{ka:210-niemczyksolano-natphys-ultrastrong}%
  \BibitemOpen
  \bibfield  {author} {\bibinfo {author} {\bibfnamefont {T.}~\bibnamefont
  {Niemczyk}}, \bibinfo {author} {\bibfnamefont {F.}~\bibnamefont {Deppe}},
  \bibinfo {author} {\bibfnamefont {H.}~\bibnamefont {Huebl}}, \bibinfo
  {author} {\bibfnamefont {E.~P.}\ \bibnamefont {Menzel}}, \bibinfo {author}
  {\bibfnamefont {F.}~\bibnamefont {Hocke}}, \bibinfo {author} {\bibfnamefont
  {M.~J.}\ \bibnamefont {Schwarz}}, \bibinfo {author} {\bibfnamefont {J.~J.}\
  \bibnamefont {Garcia-Ripoll}}, \bibinfo {author} {\bibfnamefont
  {D.}~\bibnamefont {Zueco}}, \bibinfo {author} {\bibfnamefont
  {T.}~\bibnamefont {Hummer}}, \bibinfo {author} {\bibfnamefont
  {E.}~\bibnamefont {Solano}}, \bibinfo {author} {\bibfnamefont
  {A.}~\bibnamefont {Marx}}, \ and\ \bibinfo {author} {\bibfnamefont
  {R.}~\bibnamefont {Gross}},\ }\href {\doibase 10.1038/nphys1730} {\bibfield
  {journal} {\bibinfo  {journal} {Nat. Phys.}\ }\textbf {\bibinfo {volume} {6}},\
  \bibinfo {pages} {772} (\bibinfo {year} {2010})}\BibitemShut {NoStop}%
\bibitem [{\citenamefont {Nakamura}\ and\ \citenamefont
  {Yamamoto}(2013)}]{ka:213-nakamurayam-ieee-microwphot}%
  \BibitemOpen
  \bibfield  {author} {\bibinfo {author} {\bibfnamefont {Y.}~\bibnamefont
  {Nakamura}}\ and\ \bibinfo {author} {\bibfnamefont {T.}~\bibnamefont
  {Yamamoto}},\ }\href {\doibase 10.1109/JPHOT.2013.2252005} {\bibfield
  {journal} {\bibinfo  {journal} {IEEE Photonics Journal}\ }\textbf {\bibinfo
  {volume} {5}} (\bibinfo {year} {2013}),\
  10.1109/JPHOT.2013.2252005}\BibitemShut {NoStop}%
\bibitem [{\citenamefont {Paladino}\ \emph {et~al.}(2014)\citenamefont
  {Paladino}, \citenamefont {Galperin}, \citenamefont {Falci},\ and\
  \citenamefont {Altshuler}}]{kr:214-paladino-rmp}%
  \BibitemOpen
  \bibfield  {author} {\bibinfo {author} {\bibfnamefont {E.}~\bibnamefont
  {Paladino}}, \bibinfo {author} {\bibfnamefont {Y.}~\bibnamefont {Galperin}},
  \bibinfo {author} {\bibfnamefont {G.}~\bibnamefont {Falci}}, \ and\ \bibinfo
  {author} {\bibfnamefont {B.}~\bibnamefont {Altshuler}},\ }\href {\doibase
  10.1103/RevModPhys.86.361} {\bibfield  {journal} {\bibinfo  {journal} {Rev.
  Mod. Phys.}\ }\textbf {\bibinfo {volume} {86}},\ \bibinfo {pages} {361}
  (\bibinfo {year} {2014})}\BibitemShut {NoStop}%
\bibitem [{\citenamefont {Bylander}\ \emph {et~al.}(2011)\citenamefont
  {Bylander}, \citenamefont {Gustavsson}, \citenamefont {Yan}, \citenamefont
  {Yoshihara}, \citenamefont {Harrabi}, \citenamefont {Fitch}, \citenamefont
  {Cory}, \citenamefont {Nakamura}, \citenamefont {Tsai},\ and\ \citenamefont
  {Oliver}}]{ka:211-bylander-natphys}%
  \BibitemOpen
  \bibfield  {author} {\bibinfo {author} {\bibfnamefont {J.}~\bibnamefont
  {Bylander}}, \bibinfo {author} {\bibfnamefont {S.}~\bibnamefont
  {Gustavsson}}, \bibinfo {author} {\bibfnamefont {F.}~\bibnamefont {Yan}},
  \bibinfo {author} {\bibfnamefont {F.}~\bibnamefont {Yoshihara}}, \bibinfo
  {author} {\bibfnamefont {K.}~\bibnamefont {Harrabi}}, \bibinfo {author}
  {\bibfnamefont {G.}~\bibnamefont {Fitch}}, \bibinfo {author} {\bibfnamefont
  {D.~G.}\ \bibnamefont {Cory}}, \bibinfo {author} {\bibfnamefont
  {Y.}~\bibnamefont {Nakamura}}, \bibinfo {author} {\bibfnamefont {J.-S.}\
  \bibnamefont {Tsai}}, \ and\ \bibinfo {author} {\bibfnamefont {W.~D.}\
  \bibnamefont {Oliver}},\ }\href {\doibase doi:10.1038/nphys1994} {\bibfield
  {journal} {\bibinfo  {journal} {Nat. Phys.}\ }\textbf {\bibinfo {volume}
  {7}},\ \bibinfo {pages} {565} (\bibinfo {year} {2011})}\BibitemShut {NoStop}%
\bibitem [{\citenamefont {Rigetti}\ \emph {et~al.}(2012)\citenamefont
  {Rigetti}, \citenamefont {Gambetta}, \citenamefont {Poletto}, \citenamefont
  {Plourde}, \citenamefont {Chow}, \citenamefont {C\'orcoles}, \citenamefont
  {Smolin}, \citenamefont {Merkel}, \citenamefont {Rozen}, \citenamefont
  {Keefe}, \citenamefont {Rothwell}, \citenamefont {Ketchen},\ and\
  \citenamefont {Steffen}}]{ka:212-rigettisteffen-prb-trasmonshapphire}%
  \BibitemOpen
  \bibfield  {author} {\bibinfo {author} {\bibfnamefont {C.}~\bibnamefont
  {Rigetti}}, \bibinfo {author} {\bibfnamefont {J.~M.}\ \bibnamefont
  {Gambetta}}, \bibinfo {author} {\bibfnamefont {S.}~\bibnamefont {Poletto}},
  \bibinfo {author} {\bibfnamefont {B.~L.~T.}\ \bibnamefont {Plourde}},
  \bibinfo {author} {\bibfnamefont {J.~M.}\ \bibnamefont {Chow}}, \bibinfo
  {author} {\bibfnamefont {A.~D.}\ \bibnamefont {C\'orcoles}}, \bibinfo
  {author} {\bibfnamefont {J.~A.}\ \bibnamefont {Smolin}}, \bibinfo {author}
  {\bibfnamefont {S.~T.}\ \bibnamefont {Merkel}}, \bibinfo {author}
  {\bibfnamefont {J.~R.}\ \bibnamefont {Rozen}}, \bibinfo {author}
  {\bibfnamefont {G.~A.}\ \bibnamefont {Keefe}}, \bibinfo {author}
  {\bibfnamefont {M.~B.}\ \bibnamefont {Rothwell}}, \bibinfo {author}
  {\bibfnamefont {M.~B.}\ \bibnamefont {Ketchen}}, \ and\ \bibinfo {author}
  {\bibfnamefont {M.}~\bibnamefont {Steffen}},\ }\href {\doibase
  10.1103/PhysRevB.86.100506} {\bibfield  {journal} {\bibinfo  {journal} {Phys.
  Rev. B}\ }\textbf {\bibinfo {volume} {86}},\ \bibinfo {pages} {100506}
  (\bibinfo {year} {2012})}\BibitemShut {NoStop}%
\bibitem [{\citenamefont {Stern}\ \emph {et~al.}(2014)\citenamefont {Stern},
  \citenamefont {Catelani}, \citenamefont {Kubo}, \citenamefont {Grezes},
  \citenamefont {Bienfait}, \citenamefont {Vion}, \citenamefont {Esteve},\ and\
  \citenamefont {Bertet}}]{ka:214-sternsaclay-prl-fluxqubit3D}%
  \BibitemOpen
  \bibfield  {author} {\bibinfo {author} {\bibfnamefont {M.}~\bibnamefont
  {Stern}}, \bibinfo {author} {\bibfnamefont {G.}~\bibnamefont {Catelani}},
  \bibinfo {author} {\bibfnamefont {Y.}~\bibnamefont {Kubo}}, \bibinfo {author}
  {\bibfnamefont {C.}~\bibnamefont {Grezes}}, \bibinfo {author} {\bibfnamefont
  {A.}~\bibnamefont {Bienfait}}, \bibinfo {author} {\bibfnamefont
  {D.}~\bibnamefont {Vion}}, \bibinfo {author} {\bibfnamefont {D.}~\bibnamefont
  {Esteve}}, \ and\ \bibinfo {author} {\bibfnamefont {P.}~\bibnamefont
  {Bertet}},\ }\href {\doibase 10.1103/PhysRevLett.113.123601} {\bibfield
  {journal} {\bibinfo  {journal} {Phys. Rev. Lett.}\ }\textbf {\bibinfo
  {volume} {113}},\ \bibinfo {pages} {123601} (\bibinfo {year}
  {2014})}\BibitemShut {NoStop}%
\bibitem [{\citenamefont {Liu}\ \emph {et~al.}(2005)\citenamefont {Liu},
  \citenamefont {You}, \citenamefont {Wei}, \citenamefont {Sun},\ and\
  \citenamefont {Nori}}]{ka:205-liunori-prl-adiabaticpassage}%
  \BibitemOpen
  \bibfield  {author} {\bibinfo {author} {\bibfnamefont {Y.-x.}\ \bibnamefont
  {Liu}}, \bibinfo {author} {\bibfnamefont {J.~Q.}\ \bibnamefont {You}},
  \bibinfo {author} {\bibfnamefont {L.~F.}\ \bibnamefont {Wei}}, \bibinfo
  {author} {\bibfnamefont {C.~P.}\ \bibnamefont {Sun}}, \ and\ \bibinfo
  {author} {\bibfnamefont {F.}~\bibnamefont {Nori}},\ }\href {\doibase
  10.1103/PhysRevLett.95.087001} {\bibfield  {journal} {\bibinfo  {journal}
  {Phys. Rev. Lett.}\ }\textbf {\bibinfo {volume} {95}},\ \bibinfo {pages}
  {087001} (\bibinfo {year} {2005})}\BibitemShut {NoStop}%
\bibitem [{\citenamefont {Mariantoni}\ \emph {et~al.}(2005)\citenamefont
  {Mariantoni}, \citenamefont {Storcz}, \citenamefont {Wilhelm}, \citenamefont
  {Oliver}, \citenamefont {Emmert}, \citenamefont {Marx}, \citenamefont
  {Gross}, \citenamefont {Christ},\ and\ \citenamefont
  {Solano}}]{ku:205-mariantoni-arXiv-microwfock}%
  \BibitemOpen
  \bibfield  {author} {\bibinfo {author} {\bibfnamefont {M.}~\bibnamefont
  {Mariantoni}}, \bibinfo {author} {\bibfnamefont {M.~J.}\ \bibnamefont
  {Storcz}}, \bibinfo {author} {\bibfnamefont {F.~K.}\ \bibnamefont {Wilhelm}},
  \bibinfo {author} {\bibfnamefont {W.~D.}\ \bibnamefont {Oliver}}, \bibinfo
  {author} {\bibfnamefont {A.}~\bibnamefont {Emmert}}, \bibinfo {author}
  {\bibfnamefont {A.}~\bibnamefont {Marx}}, \bibinfo {author} {\bibfnamefont
  {R.}~\bibnamefont {Gross}}, \bibinfo {author} {\bibfnamefont
  {H.}~\bibnamefont {Christ}}, \ and\ \bibinfo {author} {\bibfnamefont
  {E.}~\bibnamefont {Solano}},\ } \bibinfo {note}
  {arXiv:cond-mat/0509737v2}\BibitemShut {NoStop}%
\bibitem [{\citenamefont {Siewert}\ \emph {et~al.}(2006)\citenamefont
  {Siewert}, \citenamefont {Brandes},\ and\ \citenamefont
  {Falci}}]{ka:206-siebrafalci-optcomm-stirap}%
  \BibitemOpen
  \bibfield  {author} {\bibinfo {author} {\bibfnamefont {J.}~\bibnamefont
  {Siewert}}, \bibinfo {author} {\bibfnamefont {T.}~\bibnamefont {Brandes}}, \
  and\ \bibinfo {author} {\bibfnamefont {G.}~\bibnamefont {Falci}},\ }\href
  {\doibase http://dx.doi.org/10.1016/j.optcom.2005.12.083} {\bibfield
  {journal} {\bibinfo  {journal} {Opt. Commun.}\ }\textbf {\bibinfo {volume}
  {264}},\ \bibinfo {pages} {435-440 } (\bibinfo {year} {2006})}
  %,\ \bibinfo {note}   {quantum Control of Light and Matter. In honor of the 70th birthday of Bruce  Shore}
  \BibitemShut {NoStop}%
\bibitem [{\citenamefont {Wei}\ \emph {et~al.}(2008)\citenamefont {Wei},
  \citenamefont {Johansson}, \citenamefont {Cen}, \citenamefont {Ashhab},\ and\
  \citenamefont {Nori}}]{ka:208-weinori-prl-stirapqcomp}%
  \BibitemOpen
  \bibfield  {author} {\bibinfo {author} {\bibfnamefont {L.~F.}\ \bibnamefont
  {Wei}}, \bibinfo {author} {\bibfnamefont {J.~R.}\ \bibnamefont {Johansson}},
  \bibinfo {author} {\bibfnamefont {L.~X.}\ \bibnamefont {Cen}}, \bibinfo
  {author} {\bibfnamefont {S.}~\bibnamefont {Ashhab}}, \ and\ \bibinfo {author}
  {\bibfnamefont {F.}~\bibnamefont {Nori}},\ }\href {\doibase
  10.1103/PhysRevLett.100.113601} {\bibfield  {journal} {\bibinfo  {journal}
  {Phys. Rev. Lett.}\ }\textbf {\bibinfo {volume} {100}},\ \bibinfo {pages}
  {113601} (\bibinfo {year} {2008})}\BibitemShut {NoStop}%
\bibitem [{\citenamefont {Siewert}\ \emph {et~al.}(2009)\citenamefont
  {Siewert}, \citenamefont {Brandes},\ and\ \citenamefont
  {Falci}}]{ka:209-siebrafalci-prb}%
  \BibitemOpen
  \bibfield  {author} {\bibinfo {author} {\bibfnamefont {J.}~\bibnamefont
  {Siewert}}, \bibinfo {author} {\bibfnamefont {T.}~\bibnamefont {Brandes}}, \
  and\ \bibinfo {author} {\bibfnamefont {G.}~\bibnamefont {Falci}},\ }\href
  {\doibase 10.1103/PhysRevB.79.024504} {\bibfield  {journal} {\bibinfo
  {journal} {Phys. Rev. B}\ }\textbf {\bibinfo {volume} {79}},\ \bibinfo
  {pages} {024504} (\bibinfo {year} {2009})}\BibitemShut {NoStop}%
\bibitem [{\citenamefont {Wilk}\ \emph {et~al.}(2007)\citenamefont {Wilk},
  \citenamefont {Webster}, \citenamefont {Kuhn},\ and\ \citenamefont
  {Rempe}}]{ka:207-wilkrempe-science-singleatomph}%
  \BibitemOpen
  \bibfield  {author} {\bibinfo {author} {\bibfnamefont {T.}~\bibnamefont
  {Wilk}}, \bibinfo {author} {\bibfnamefont {S.~C.}\ \bibnamefont {Webster}},
  \bibinfo {author} {\bibfnamefont {A.}~\bibnamefont {Kuhn}}, \ and\ \bibinfo
  {author} {\bibfnamefont {G.}~\bibnamefont {Rempe}},\ }\href {\doibase
  10.1126/science.1143835} {\bibfield  {journal} {\bibinfo  {journal}
  {Science}\ }\textbf {\bibinfo {volume} {317}},\ \bibinfo {pages} {488}
  (\bibinfo {year} {2007})}\BibitemShut {NoStop}%
\bibitem [{\citenamefont {Bergmann}\ \emph {et~al.}(2015)\citenamefont
  {Bergmann}, \citenamefont {Vitanov},\ and\ \citenamefont
  {Shore}}]{kr:215-bergmannetal-jchemphys-revstirap}%
  \BibitemOpen
  \bibfield  {author} {\bibinfo {author} {\bibfnamefont {K.}~\bibnamefont
  {Bergmann}}, \bibinfo {author} {\bibfnamefont {N.~V.}\ \bibnamefont
  {Vitanov}}, \ and\ \bibinfo {author} {\bibfnamefont {B.~W.}\ \bibnamefont
  {Shore}},\ }\href {\doibase 10.1063/1.4916903} {\bibfield  {journal}
  {\bibinfo  {journal} {Jour. Chem. Phys.}\ }\textbf {\bibinfo {volume}
  {142}},\ \bibinfo {pages} {170901} (\bibinfo {year} {2015})}\BibitemShut
  {NoStop}%
\bibitem [{\citenamefont {Di~Stefano}\ \emph {et~al.}(2015)\citenamefont
  {Di~Stefano}, \citenamefont {Paladino}, \citenamefont {D'Arrigo},\ and\
  \citenamefont {Falci}}]{ka:215-distefano}%
  \BibitemOpen
  \bibfield  {author} {\bibinfo {author} {\bibfnamefont {P.~G.}\ \bibnamefont
  {Di~Stefano}}, \bibinfo {author} {\bibfnamefont {E.}~\bibnamefont
  {Paladino}}, \bibinfo {author} {\bibfnamefont {A.}~\bibnamefont {D'Arrigo}},
  \ and\ \bibinfo {author} {\bibfnamefont {G.}~\bibnamefont {Falci}},\ }\href
  {\doibase 10.1103/PhysRevB.91.224506} {\bibfield  {journal} {\bibinfo
  {journal} {Phys. Rev. B}\ }\textbf {\bibinfo {volume} {91}},\ \bibinfo
  {pages} {224506} (\bibinfo {year} {2015})}\BibitemShut {NoStop}%
\bibitem [{\citenamefont {M\"oller}\ \emph {et~al.}(2008)\citenamefont
  {M\"oller}, \citenamefont {Madsen},\ and\ \citenamefont
  {M\"olmer}}]{ka:208-dmoller-prl-adiabgates}%
  \BibitemOpen
  \bibfield  {author} {\bibinfo {author} {\bibfnamefont {D.}~\bibnamefont
  {M\"oller}}, \bibinfo {author} {\bibfnamefont {L.~B.}\ \bibnamefont
  {Madsen}}, \ and\ \bibinfo {author} {\bibfnamefont {K.}~\bibnamefont
  {M\"olmer}},\ }\href {\doibase 10.1103/PhysRevLett.100.170504} {\bibfield
  {journal} {\bibinfo  {journal} {Phys. Rev. Lett.}\ }\textbf {\bibinfo
  {volume} {100}},\ \bibinfo {pages} {170504} (\bibinfo {year}
  {2008})}\BibitemShut {NoStop}%
\bibitem [{\citenamefont {Yang}\ \emph {et~al.}(2004)\citenamefont {Yang},
  \citenamefont {Chu},\ and\ \citenamefont {Han}}]{ka:204-yang-prl-fluxstirap}%
  \BibitemOpen
  \bibfield  {author} {\bibinfo {author} {\bibfnamefont {C.-P.}\ \bibnamefont
  {Yang}}, \bibinfo {author} {\bibfnamefont {S.-I.}\ \bibnamefont {Chu}}, \
  and\ \bibinfo {author} {\bibfnamefont {S.}~\bibnamefont {Han}},\ }\href
  {\doibase 10.1103/PhysRevLett.92.117902} {\bibfield  {journal} {\bibinfo
  {journal} {Phys. Rev. Lett.}\ }\textbf {\bibinfo {volume} {92}},\ \bibinfo
  {pages} {117902} (\bibinfo {year} {2004})}\BibitemShut {NoStop}%
\bibitem [{\citenamefont {Kis}\ and\ \citenamefont
  {Paspalakis}(2004)}]{ka:204-kis-prb-fluxstirap}%
  \BibitemOpen
  \bibfield  {author} {\bibinfo {author} {\bibfnamefont {Z.}~\bibnamefont
  {Kis}}\ and\ \bibinfo {author} {\bibfnamefont {E.}~\bibnamefont
  {Paspalakis}},\ }\href {\doibase 10.1103/PhysRevB.69.024510} {\bibfield
  {journal} {\bibinfo  {journal} {Phys. Rev. B}\ }\textbf {\bibinfo {volume}
  {69}},\ \bibinfo {pages} {024510} (\bibinfo {year} {2004})}\BibitemShut
  {NoStop}%
\bibitem [{\citenamefont {Kr\'al}\ \emph {et~al.}(2007)\citenamefont {Kr\'al},
  \citenamefont {Thanopulos},\ and\ \citenamefont
  {Shapiro}}]{kr:207-kral-rmp-controladpass}%
  \BibitemOpen
  \bibfield  {author} {\bibinfo {author} {\bibfnamefont {P.}~\bibnamefont
  {Kr\'al}}, \bibinfo {author} {\bibfnamefont {I.}~\bibnamefont {Thanopulos}},
  \ and\ \bibinfo {author} {\bibfnamefont {M.}~\bibnamefont {Shapiro}},\ }\href
  {\doibase 10.1103/RevModPhys.79.53} {\bibfield  {journal} {\bibinfo
  {journal} {Rev. Mod. Phys.}\ }\textbf {\bibinfo {volume} {79}},\ \bibinfo
  {pages} {53} (\bibinfo {year} {2007})}\BibitemShut {NoStop}%
\bibitem [{\citenamefont {Vitanov}\ \emph {et~al.}(2001)\citenamefont
  {Vitanov}, \citenamefont {Fleischhauer}, \citenamefont {Shore},\ and\
  \citenamefont {Bergmann}}]{kr:201-vitanov-advatmolopt}%
  \BibitemOpen
  \bibfield  {author} {\bibinfo {author} {\bibfnamefont {N.}~\bibnamefont
  {Vitanov}}, \bibinfo {author} {\bibfnamefont {M.}~\bibnamefont
  {Fleischhauer}}, \bibinfo {author} {\bibfnamefont {B.}~\bibnamefont {Shore}},
  \ and\ \bibinfo {author} {\bibfnamefont {K.}~\bibnamefont {Bergmann}},\
  }\href {\doibase DOI: 10.1016/S1049-250X(01)80063-X} {\bibfield  {journal}
  {\bibinfo  {journal} {Adv. in At. Mol. and Opt. Phys.}\ }\textbf {\bibinfo
  {volume} {46}},\ \bibinfo {pages} {55} (\bibinfo {year} {2001})}\BibitemShut
  {NoStop}%
\bibitem [{\citenamefont {Kelly}\ \emph {et~al.}(2010)\citenamefont {Kelly},
  \citenamefont {Dutton}, \citenamefont {Schlafer}, \citenamefont {Mookerji},
  \citenamefont {Ohki}, \citenamefont {Kline},\ and\ \citenamefont
  {Pappas}}]{ka:210-kellypappas-prl-cpt}%
  \BibitemOpen
  \bibfield  {author} {\bibinfo {author} {\bibfnamefont {W.~R.}\ \bibnamefont
  {Kelly}}, \bibinfo {author} {\bibfnamefont {Z.}~\bibnamefont {Dutton}},
  \bibinfo {author} {\bibfnamefont {J.}~\bibnamefont {Schlafer}}, \bibinfo
  {author} {\bibfnamefont {B.}~\bibnamefont {Mookerji}}, \bibinfo {author}
  {\bibfnamefont {T.~A.}\ \bibnamefont {Ohki}}, \bibinfo {author}
  {\bibfnamefont {J.~S.}\ \bibnamefont {Kline}}, \ and\ \bibinfo {author}
  {\bibfnamefont {D.~P.}\ \bibnamefont {Pappas}},\ }\href {\doibase
  10.1103/PhysRevLett.104.163601} {\bibfield  {journal} {\bibinfo  {journal}
  {Phys. Rev. Lett.}\ }\textbf {\bibinfo {volume} {104}},\ \bibinfo {pages}
  {163601} (\bibinfo {year} {2010})}\BibitemShut {NoStop}%
\bibitem [{\citenamefont {Bergmann}\ \emph {et~al.}(1998)\citenamefont
  {Bergmann}, \citenamefont {Theuer},\ and\ \citenamefont
  {Shore}}]{kr:198-bergmann-rmp-stirap}%
  \BibitemOpen
  \bibfield  {author} {\bibinfo {author} {\bibfnamefont {K.}~\bibnamefont
  {Bergmann}}, \bibinfo {author} {\bibfnamefont {H.}~\bibnamefont {Theuer}}, \
  and\ \bibinfo {author} {\bibfnamefont {B.}~\bibnamefont {Shore}},\ }\href
  {\doibase 10.1103/RevModPhys.70.1003} {\bibfield  {journal} {\bibinfo
  {journal} {Rev. Mod. Phys.}\ }\textbf {\bibinfo {volume} {70}},\ \bibinfo
  {pages} {1003} (\bibinfo {year} {1998})}\BibitemShut {NoStop}%
\bibitem [{\citenamefont {Xiang}\ \emph {et~al.}(2013)\citenamefont {Xiang},
  \citenamefont {Ashhab}, \citenamefont {You},\ and\ \citenamefont
  {Nori}}]{kr:213-xiangnori-rmp-hybrid}%
  \BibitemOpen
  \bibfield  {author} {\bibinfo {author} {\bibfnamefont {Z.-L.}\ \bibnamefont
  {Xiang}}, \bibinfo {author} {\bibfnamefont {S.}~\bibnamefont {Ashhab}},
  \bibinfo {author} {\bibfnamefont {J.}~\bibnamefont {You}}, \ and\ \bibinfo
  {author} {\bibfnamefont {F.}~\bibnamefont {Nori}},\ }\href
  {http://dx.doi.org/10.1103/RevModPhys.85.623} {\bibfield  {journal} {\bibinfo
   {journal} {Rev. Mod. Phys.}\ }\textbf {\bibinfo {volume} {85}},\ \bibinfo
  {pages} {623} (\bibinfo {year} {2013})}\BibitemShut {NoStop}%
\bibitem [{\citenamefont {Vion}\ \emph {et~al.}(2002)\citenamefont {Vion},
  \citenamefont {Aassime}, \citenamefont {Cottet}, \citenamefont {Joyez},
  \citenamefont {Pothier}, \citenamefont {Urbina}, \citenamefont {Esteve},\
  and\ \citenamefont {Devoret}}]{ka:202-vion-science}%
  \BibitemOpen
  \bibfield  {author} {\bibinfo {author} {\bibfnamefont {D.}~\bibnamefont
  {Vion}}, \bibinfo {author} {\bibfnamefont {A.}~\bibnamefont {Aassime}},
  \bibinfo {author} {\bibfnamefont {A.}~\bibnamefont {Cottet}}, \bibinfo
  {author} {\bibfnamefont {P.}~\bibnamefont {Joyez}}, \bibinfo {author}
  {\bibfnamefont {H.}~\bibnamefont {Pothier}}, \bibinfo {author} {\bibfnamefont
  {C.}~\bibnamefont {Urbina}}, \bibinfo {author} {\bibfnamefont
  {D.}~\bibnamefont {Esteve}}, \ and\ \bibinfo {author} {\bibfnamefont {M.~H.}\
  \bibnamefont {Devoret}},\ }\href {\doibase 10.1126/science.1069372}
  {\bibfield  {journal} {\bibinfo  {journal} {Science}\ }\textbf {\bibinfo
  {volume} {296}},\ \bibinfo {pages} {886} (\bibinfo {year}
  {2002})}\BibitemShut {NoStop}%
\bibitem [{\citenamefont {Falci}\ \emph {et~al.}(2005)\citenamefont {Falci},
  \citenamefont {D'Arrigo}, \citenamefont {Mastellone},\ and\ \citenamefont
  {Paladino}}]{ka:205-falci-prl}%
  \BibitemOpen
  \bibfield  {author} {\bibinfo {author} {\bibfnamefont {G.}~\bibnamefont
  {Falci}}, \bibinfo {author} {\bibfnamefont {A.}~\bibnamefont {D'Arrigo}},
  \bibinfo {author} {\bibfnamefont {A.}~\bibnamefont {Mastellone}}, \ and\
  \bibinfo {author} {\bibfnamefont {E.}~\bibnamefont {Paladino}},\ }\href
  {\doibase 10.1103/PhysRevLett.94.167002} {\bibfield  {journal} {\bibinfo
  {journal} {Phys. Rev. Lett.}\ }\textbf {\bibinfo {volume} {94}},\ \bibinfo
  {pages} {167002} (\bibinfo {year} {2005})}\BibitemShut {NoStop}%
\bibitem [{not()}]{note:suppl-mat}%
  \BibitemOpen
  \href@noop {} {}\bibinfo {edition} {See supplemental material at [url will be
  inserted by editor] for further details on magnus expansion, superconducting
  devices and models for dynamics and noise.}\BibitemShut {Stop}%
\bibitem [{\citenamefont {Falci}\ \emph {et~al.}(2013)\citenamefont {Falci},
  \citenamefont {La~Cognata}, \citenamefont {Berritta}, \citenamefont
  {D'Arrigo}, \citenamefont {Paladino},\ and\ \citenamefont
  {Spagnolo}}]{ka:213-falci-prb-stirapcpb}%
  \BibitemOpen
  \bibfield  {author} {\bibinfo {author} {\bibfnamefont {G.}~\bibnamefont
  {Falci}}, \bibinfo {author} {\bibfnamefont {A.}~\bibnamefont {La~Cognata}},
  \bibinfo {author} {\bibfnamefont {M.}~\bibnamefont {Berritta}}, \bibinfo
  {author} {\bibfnamefont {A.}~\bibnamefont {D'Arrigo}}, \bibinfo {author}
  {\bibfnamefont {E.}~\bibnamefont {Paladino}}, \ and\ \bibinfo {author}
  {\bibfnamefont {B.}~\bibnamefont {Spagnolo}},\ }\href {\doibase
  10.1103/PhysRevB.87.214515} {\bibfield  {journal} {\bibinfo  {journal} {Phys.
  Rev. B}\ }\textbf {\bibinfo {volume} {87}},\ \bibinfo {pages} {214515}
  (\bibinfo {year} {2013})}\BibitemShut {NoStop}%
\bibitem [{\citenamefont {Yatsenko}\ \emph {et~al.}(1998)\citenamefont
  {Yatsenko}, \citenamefont {Gu\'erin}, \citenamefont {Halfmann}, \citenamefont
  {B\"ohmer}, \citenamefont {Shore},\ and\ \citenamefont
  {Bergmann}}]{ka:198-yatsenko-pra-stirap21th1}%
  \BibitemOpen
  \bibfield  {author} {\bibinfo {author} {\bibfnamefont {L.~P.}\ \bibnamefont
  {Yatsenko}}, \bibinfo {author} {\bibfnamefont {S.}~\bibnamefont {Gu\'erin}},
  \bibinfo {author} {\bibfnamefont {T.}~\bibnamefont {Halfmann}}, \bibinfo
  {author} {\bibfnamefont {K.}~\bibnamefont {B\"ohmer}}, \bibinfo {author}
  {\bibfnamefont {B.~W.}\ \bibnamefont {Shore}}, \ and\ \bibinfo {author}
  {\bibfnamefont {K.}~\bibnamefont {Bergmann}},\ }\href {\doibase
  10.1103/PhysRevA.58.4683} {\bibfield  {journal} {\bibinfo  {journal} {Phys.
  Rev. A}\ }\textbf {\bibinfo {volume} {58}},\ \bibinfo {pages} {4683}
  (\bibinfo {year} {1998})}\BibitemShut {NoStop}%
\bibitem [{\citenamefont {Gu\'erin}\ \emph {et~al.}(1998)\citenamefont
  {Gu\'erin}, \citenamefont {Yatsenko}, \citenamefont {Halfmann}, \citenamefont
  {Shore},\ and\ \citenamefont
  {Bergmann}}]{ka:198-guerin-pra-stirap21staticcomp}%
  \BibitemOpen
  \bibfield  {author} {\bibinfo {author} {\bibfnamefont {S.}~\bibnamefont
  {Gu\'erin}}, \bibinfo {author} {\bibfnamefont {L.~P.}\ \bibnamefont
  {Yatsenko}}, \bibinfo {author} {\bibfnamefont {T.}~\bibnamefont {Halfmann}},
  \bibinfo {author} {\bibfnamefont {B.~W.}\ \bibnamefont {Shore}}, \ and\
  \bibinfo {author} {\bibfnamefont {K.}~\bibnamefont {Bergmann}},\ }\href
  {\doibase 10.1103/PhysRevA.58.4691} {\bibfield  {journal} {\bibinfo
  {journal} {Phys. Rev. A}\ }\textbf {\bibinfo {volume} {58}},\ \bibinfo
  {pages} {4691} (\bibinfo {year} {1998})}\BibitemShut {NoStop}%
\bibitem [{\citenamefont {B\"ohmer}\ \emph {et~al.}(2001)\citenamefont
  {B\"ohmer}, \citenamefont {Halfmann}, \citenamefont {Yatsenko}, \citenamefont
  {Shore},\ and\ \citenamefont {Bergmann}}]{ka:198-bohmer-pra-stirap21exp}%
  \BibitemOpen
  \bibfield  {author} {\bibinfo {author} {\bibfnamefont {K.}~\bibnamefont
  {B\"ohmer}}, \bibinfo {author} {\bibfnamefont {T.}~\bibnamefont {Halfmann}},
  \bibinfo {author} {\bibfnamefont {L.~P.}\ \bibnamefont {Yatsenko}}, \bibinfo
  {author} {\bibfnamefont {B.~W.}\ \bibnamefont {Shore}}, \ and\ \bibinfo
  {author} {\bibfnamefont {K.}~\bibnamefont {Bergmann}},\ }\href {\doibase
  10.1103/PhysRevA.64.023404} {\bibfield  {journal} {\bibinfo  {journal} {Phys.
  Rev. A}\ }\textbf {\bibinfo {volume} {64}},\ \bibinfo {pages} {023404}
  (\bibinfo {year} {2001})}\BibitemShut {NoStop}%
\bibitem [{\citenamefont {van~der Wal}\ \emph {et~al.}(2000)\citenamefont
  {van~der Wal}, \citenamefont {ter Haar}, \citenamefont {Wilhelm},
  \citenamefont {Schouten}, \citenamefont {Harmans}, \citenamefont {Orlando},
  \citenamefont {Lloyd},\ and\ \citenamefont
  {Mooij}}]{ka:200-walmooij-science-superposition}%
  \BibitemOpen
  \bibfield  {author} {\bibinfo {author} {\bibfnamefont {C.}~\bibnamefont
  {van~der Wal}}, \bibinfo {author} {\bibfnamefont {A.}~\bibnamefont {ter
  Haar}}, \bibinfo {author} {\bibfnamefont {F.}~\bibnamefont {Wilhelm}},
  \bibinfo {author} {\bibfnamefont {R.}~\bibnamefont {Schouten}}, \bibinfo
  {author} {\bibfnamefont {C.}~\bibnamefont {Harmans}}, \bibinfo {author}
  {\bibfnamefont {T.}~\bibnamefont {Orlando}}, \bibinfo {author} {\bibfnamefont
  {S.}~\bibnamefont {Lloyd}}, \ and\ \bibinfo {author} {\bibfnamefont
  {J.}~\bibnamefont {Mooij}},\ }\href {\doibase 10.1126/science.290.5492.773}
  {\bibfield  {journal} {\bibinfo  {journal} {Science}\ }\textbf {\bibinfo
  {volume} {290}},\ \bibinfo {pages} {773} (\bibinfo {year}
  {2000})}\BibitemShut {NoStop}%
\bibitem [{\citenamefont {Wallraff}\ \emph {et~al.}(2004)\citenamefont
  {Wallraff}, \citenamefont {Schuster}, \citenamefont {Blais}, \citenamefont
  {Frunzio}, \citenamefont {Huang}, \citenamefont {Majer}, \citenamefont
  {Kumar}, \citenamefont {Girvin},\ and\ \citenamefont
  {Schoelkopf}}]{ka:204-wallraff-superqubit}%
  \BibitemOpen
  \bibfield  {author} {\bibinfo {author} {\bibfnamefont {A.}~\bibnamefont
  {Wallraff}}, \bibinfo {author} {\bibfnamefont {D.~I.}\ \bibnamefont
  {Schuster}}, \bibinfo {author} {\bibfnamefont {A.}~\bibnamefont {Blais}},
  \bibinfo {author} {\bibfnamefont {L.}~\bibnamefont {Frunzio}}, \bibinfo
  {author} {\bibfnamefont {R.-S.}\ \bibnamefont {Huang}}, \bibinfo {author}
  {\bibfnamefont {J.}~\bibnamefont {Majer}}, \bibinfo {author} {\bibfnamefont
  {S.}~\bibnamefont {Kumar}}, \bibinfo {author} {\bibfnamefont
  {S.}~\bibnamefont {Girvin}}, \ and\ \bibinfo {author} {\bibfnamefont
  {R.}~\bibnamefont {Schoelkopf}},\ }\href {\doibase 10.1038/nature02851}
  {\bibfield  {journal} {\bibinfo  {journal} {Nature}\ }\textbf {\bibinfo
  {volume} {421}},\ \bibinfo {pages} {162} (\bibinfo {year}
  {2004})}\BibitemShut {NoStop}%
\bibitem [{\citenamefont {Koch}\ \emph {et~al.}(2007)\citenamefont {Koch},
  \citenamefont {Yu}, \citenamefont {Gambetta}, \citenamefont {Houck},
  \citenamefont {Schuster}, \citenamefont {Majer}, \citenamefont {Blais},
  \citenamefont {Devoret}, \citenamefont {Girvin},\ and\ \citenamefont
  {Schoelkopf}}]{ka:207-koch-pra-transmon}%
  \BibitemOpen
  \bibfield  {author} {\bibinfo {author} {\bibfnamefont {J.}~\bibnamefont
  {Koch}}, \bibinfo {author} {\bibfnamefont {T.}~\bibnamefont {Yu}}, \bibinfo
  {author} {\bibfnamefont {J.}~\bibnamefont {Gambetta}}, \bibinfo {author}
  {\bibfnamefont {A.}~\bibnamefont {Houck}}, \bibinfo {author} {\bibfnamefont
  {D.}~\bibnamefont {Schuster}}, \bibinfo {author} {\bibfnamefont
  {J.}~\bibnamefont {Majer}}, \bibinfo {author} {\bibfnamefont
  {A.}~\bibnamefont {Blais}}, \bibinfo {author} {\bibfnamefont
  {M.}~\bibnamefont {Devoret}}, \bibinfo {author} {\bibfnamefont
  {S.}~\bibnamefont {Girvin}}, \ and\ \bibinfo {author} {\bibfnamefont
  {R.}~\bibnamefont {Schoelkopf}},\ }\href {\doibase
  10.1103/PhysRevA.76.042319} {\bibfield  {journal} {\bibinfo  {journal} {Phys.
  Rev. A}\ }\textbf {\bibinfo {volume} {76}}, \ 042319 (\bibinfo {year} {2007})}\BibitemShut {NoStop}%
\bibitem [{\citenamefont {Waugh}\ \emph {et~al.}(1968)\citenamefont {Waugh},
  \citenamefont {Huber},\ and\ \citenamefont
  {Haeberlen}}]{ka:168-waugh-prl-average-hamiltonian}%
  \BibitemOpen
  \bibfield  {author} {\bibinfo {author} {\bibfnamefont {J.~S.}\ \bibnamefont
  {Waugh}}, \bibinfo {author} {\bibfnamefont {L.~M.}\ \bibnamefont {Huber}}, \
  and\ \bibinfo {author} {\bibfnamefont {U.}~\bibnamefont {Haeberlen}},\ }\href
  {\doibase 10.1103/PhysRevLett.20.180} {\bibfield  {journal} {\bibinfo
  {journal} {Phys. Rev. Lett.}\ }\textbf {\bibinfo {volume} {20}},\ \bibinfo
  {pages} {180} (\bibinfo {year} {1968})}\BibitemShut {NoStop}%
\bibitem [{\citenamefont {Pechal}\ \emph {et~al.}(2014)\citenamefont {Pechal},
  \citenamefont {Huthmacher}, \citenamefont {Eichler}, \citenamefont
  {Zeytinoglu}, \citenamefont {Abdumalikov~Jr.}, \citenamefont {Berger},
  \citenamefont {Wallraff},\ and\ \citenamefont
  {Filipp}}]{ka:214-pechalwallraff-prx-singlephoton}%
  \BibitemOpen
  \bibfield  {author} {\bibinfo {author} {\bibfnamefont {M.}~\bibnamefont
  {Pechal}}, \bibinfo {author} {\bibfnamefont {L.}~\bibnamefont {Huthmacher}},
  \bibinfo {author} {\bibfnamefont {C.}~\bibnamefont {Eichler}}, \bibinfo
  {author} {\bibfnamefont {S.}~\bibnamefont {Zeytinoglu}}, \bibinfo {author}
  {\bibfnamefont {A.}~\bibnamefont {Abdumalikov~Jr.}}, \bibinfo {author}
  {\bibfnamefont {S.}~\bibnamefont {Berger}}, \bibinfo {author} {\bibfnamefont
  {A.}~\bibnamefont {Wallraff}}, \ and\ \bibinfo {author} {\bibfnamefont
  {S.}~\bibnamefont {Filipp}},\ }\href {\doibase 10.1103/PhysRevX.4.041010}
  {\bibfield  {journal} {\bibinfo  {journal} {Phys. Rev. X}\ }\textbf {\bibinfo
  {volume} {4}},\ \bibinfo {pages} {041010} (\bibinfo {year}
  {2014})}\BibitemShut {NoStop}%
\bibitem [{\citenamefont {Kumar}\ \emph {et~al.}(2016)\citenamefont {Kumar},
  \citenamefont {Veps\"al\"ainen}, \citenamefont {Danilin},\ and\ \citenamefont
  {Paraoanu}}]{ka:216-kumarparaoanu-natcomm-stirap}%
  \BibitemOpen
  \bibfield  {author} {\bibinfo {author} {\bibfnamefont {K.}~\bibnamefont
  {Kumar}}, \bibinfo {author} {\bibfnamefont {A.}~\bibnamefont
  {Veps\"al\"ainen}}, \bibinfo {author} {\bibfnamefont {S.}~\bibnamefont
  {Danilin}}, \ and\ \bibinfo {author} {\bibfnamefont {G.}~\bibnamefont
  {Paraoanu}},\ }\href {\doibase 10.1038/ncomms10628} {\bibfield  {journal}
  {\bibinfo  {journal} {Nat. Comm.}\ }\textbf {\bibinfo {volume} {7}},\
  \bibinfo {pages} {10628} (\bibinfo {year} {2016})}\BibitemShut {NoStop}%
\bibitem [{\citenamefont {Xu}\ \emph {et~al.}(2016)\citenamefont {Xu},
  \citenamefont {Song}, \citenamefont {Liu}, \citenamefont {Xue}, \citenamefont
  {Su}, \citenamefont {Deng}, \citenamefont {Tian}, \citenamefont {Zheng},
  \citenamefont {Han}, \citenamefont {Zhong}, \citenamefont {Wang},
  \citenamefont {Liu},\ and\ \citenamefont
  {Zhao}}]{ka:216-xuzhao-ncomm-stirap}%
  \BibitemOpen
  \bibfield  {author} {\bibinfo {author} {\bibfnamefont {H.~K.}\ \bibnamefont
  {Xu}}, \bibinfo {author} {\bibfnamefont {C.}~\bibnamefont {Song}}, \bibinfo
  {author} {\bibfnamefont {W.~Y.}\ \bibnamefont {Liu}}, \bibinfo {author}
  {\bibfnamefont {G.~M.}\ \bibnamefont {Xue}}, \bibinfo {author} {\bibfnamefont
  {F.~F.}\ \bibnamefont {Su}}, \bibinfo {author} {\bibfnamefont
  {H.}~\bibnamefont {Deng}}, \bibinfo {author} {\bibfnamefont {Y.}~\bibnamefont
  {Tian}}, \bibinfo {author} {\bibfnamefont {D.~N.}\ \bibnamefont {Zheng}},
  \bibinfo {author} {\bibfnamefont {S.}~\bibnamefont {Han}}, \bibinfo {author}
  {\bibfnamefont {Y.~P.}\ \bibnamefont {Zhong}}, \bibinfo {author}
  {\bibfnamefont {H.}~\bibnamefont {Wang}}, \bibinfo {author} {\bibfnamefont
  {Y.-x.}\ \bibnamefont {Liu}}, \ and\ \bibinfo {author} {\bibfnamefont
  {S.~P.}\ \bibnamefont {Zhao}},\ }\href
  {http://dx.doi.org/10.1038/ncomms11018} {\bibfield  {journal} {\bibinfo
  {journal} {Nat. Commun.}\ }\textbf {\bibinfo {volume} {7}},\ 11018 (\bibinfo {year}
  {2016})}\BibitemShut {NoStop}%
\end{thebibliography}

\begin{thebibliography}{9}%
\makeatletter
\providecommand \@ifxundefined [1]{%
 \@ifx{#1\undefined}
}%
\providecommand \@ifnum [1]{%
 \ifnum #1\expandafter \@firstoftwo
 \else \expandafter \@secondoftwo
 \fi
}%
\providecommand \@ifx [1]{%
 \ifx #1\expandafter \@firstoftwo
 \else \expandafter \@secondoftwo
 \fi
}%
\providecommand \natexlab [1]{#1}%
\providecommand \enquote  [1]{``#1''}%
\providecommand \bibnamefont  [1]{#1}%
\providecommand \bibfnamefont [1]{#1}%
\providecommand \citenamefont [1]{#1}%
\providecommand \href@noop [0]{\@secondoftwo}%
\providecommand \href [0]{\begingroup \@sanitize@url \@href}%
\providecommand \@href[1]{\@@startlink{#1}\@@href}%
\providecommand \@@href[1]{\endgroup#1\@@endlink}%
\providecommand \@sanitize@url [0]{\catcode `\\12\catcode `\$12\catcode
  `\&12\catcode `\#12\catcode `\^12\catcode `\_12\catcode `\%12\relax}%
\providecommand \@@startlink[1]{}%
\providecommand \@@endlink[0]{}%
\providecommand \url  [0]{\begingroup\@sanitize@url \@url }%
\providecommand \@url [1]{\endgroup\@href {#1}{\urlprefix }}%
\providecommand \urlprefix  [0]{URL }%
\providecommand \Eprint [0]{\href }%
\providecommand \doibase [0]{http://dx.doi.org/}%
\providecommand \selectlanguage [0]{\@gobble}%
\providecommand \bibinfo  [0]{\@secondoftwo}%
\providecommand \bibfield  [0]{\@secondoftwo}%
\providecommand \translation [1]{[#1]}%
\providecommand \BibitemOpen [0]{}%
\providecommand \bibitemStop [0]{}%
\providecommand \bibitemNoStop [0]{.\EOS\space}%
\providecommand \EOS [0]{\spacefactor3000\relax}%
\providecommand \BibitemShut  [1]{\csname bibitem#1\endcsname}%
\let\auto@bib@innerbib\@empty
%</preamble>
\bibitem [{\citenamefont {Stern}\ \emph {et~al.}(2014)\citenamefont {Stern},
  \citenamefont {Catelani}, \citenamefont {Kubo}, \citenamefont {Grezes},
  \citenamefont {Bienfait}, \citenamefont {Vion}, \citenamefont {Esteve},\ and\
  \citenamefont {Bertet}}]{ka:214:sternvion-prl-flux}%
  \BibitemOpen
  \bibfield  {author} {\bibinfo {author} {\bibfnamefont {M.}~\bibnamefont
  {Stern}}, \bibinfo {author} {\bibfnamefont {G.}~\bibnamefont {Catelani}},
  \bibinfo {author} {\bibfnamefont {Y.}~\bibnamefont {Kubo}}, \bibinfo {author}
  {\bibfnamefont {C.}~\bibnamefont {Grezes}}, \bibinfo {author} {\bibfnamefont
  {A.}~\bibnamefont {Bienfait}}, \bibinfo {author} {\bibfnamefont
  {D.}~\bibnamefont {Vion}}, \bibinfo {author} {\bibfnamefont {D.}~\bibnamefont
  {Esteve}}, \ and\ \bibinfo {author} {\bibfnamefont {P.}~\bibnamefont
  {Bertet}},\ }\href {\doibase 10.1103/PhysRevLett.113.123601} {\bibfield
  {journal} {\bibinfo  {journal} {Phys. Rev. Lett.}\ }\textbf {\bibinfo
  {volume} {113}},\ \bibinfo {pages} {123601} (\bibinfo {year}
  {2014})}\BibitemShut {NoStop}%
\bibitem [{\citenamefont {Bylander}\ \emph {et~al.}(2011)\citenamefont
  {Bylander}, \citenamefont {Gustavsson}, \citenamefont {Yan}, \citenamefont
  {Yoshihara}, \citenamefont {Harrabi}, \citenamefont {Fitch}, \citenamefont
  {Cory}, \citenamefont {Nakamura}, \citenamefont {Tsai},\ and\ \citenamefont
  {Oliver}}]{ka:211-bylander-natphys}%
  \BibitemOpen
  \bibfield  {author} {\bibinfo {author} {\bibfnamefont {J.}~\bibnamefont
  {Bylander}}, \bibinfo {author} {\bibfnamefont {S.}~\bibnamefont
  {Gustavsson}}, \bibinfo {author} {\bibfnamefont {F.}~\bibnamefont {Yan}},
  \bibinfo {author} {\bibfnamefont {F.}~\bibnamefont {Yoshihara}}, \bibinfo
  {author} {\bibfnamefont {K.}~\bibnamefont {Harrabi}}, \bibinfo {author}
  {\bibfnamefont {G.}~\bibnamefont {Fitch}}, \bibinfo {author} {\bibfnamefont
  {D.~G.}\ \bibnamefont {Cory}}, \bibinfo {author} {\bibfnamefont
  {Y.}~\bibnamefont {Nakamura}}, \bibinfo {author} {\bibfnamefont {J.-S.}\
  \bibnamefont {Tsai}}, \ and\ \bibinfo {author} {\bibfnamefont {W.~D.}\
  \bibnamefont {Oliver}},\ }\href {\doibase doi:10.1038/nphys1994} {\bibfield
  {journal} {\bibinfo  {journal} {Nature Physics}\ }\textbf {\bibinfo {volume}
  {7}},\ \bibinfo {pages} {565} (\bibinfo {year} {2011})}\BibitemShut {NoStop}%
\bibitem [{\citenamefont {Koch}\ \emph {et~al.}(2007)\citenamefont {Koch},
  \citenamefont {Yu}, \citenamefont {Gambetta}, \citenamefont {Houck},
  \citenamefont {Schuster}, \citenamefont {Majer}, \citenamefont {Blais},
  \citenamefont {Devoret}, \citenamefont {Girvin},\ and\ \citenamefont
  {Schoelkopf}}]{ka:207-koch-pra-transmon}%
  \BibitemOpen
  \bibfield  {author} {\bibinfo {author} {\bibfnamefont {J.}~\bibnamefont
  {Koch}}, \bibinfo {author} {\bibfnamefont {T.~M.}\ \bibnamefont {Yu}},
  \bibinfo {author} {\bibfnamefont {J.}~\bibnamefont {Gambetta}}, \bibinfo
  {author} {\bibfnamefont {A.~A.}\ \bibnamefont {Houck}}, \bibinfo {author}
  {\bibfnamefont {D.~I.}\ \bibnamefont {Schuster}}, \bibinfo {author}
  {\bibfnamefont {J.}~\bibnamefont {Majer}}, \bibinfo {author} {\bibfnamefont
  {A.}~\bibnamefont {Blais}}, \bibinfo {author} {\bibfnamefont {M.~H.}\
  \bibnamefont {Devoret}}, \bibinfo {author} {\bibfnamefont {S.~M.}\
  \bibnamefont {Girvin}}, \ and\ \bibinfo {author} {\bibfnamefont {R.~J.}\
  \bibnamefont {Schoelkopf}},\ }\href {\doibase 10.1103/PhysRevA.76.042319}
  {\bibfield  {journal} {\bibinfo  {journal} {Phys. Rev. A}\ }\textbf {\bibinfo
  {volume} {76}},\ \bibinfo {pages} {042319} (\bibinfo {year}
  {2007})}\BibitemShut {NoStop}%
\bibitem [{\citenamefont {Rigetti}\ \emph {et~al.}(2012)\citenamefont
  {Rigetti}, \citenamefont {Gambetta}, \citenamefont {Poletto}, \citenamefont
  {Plourde}, \citenamefont {Chow}, \citenamefont {C\'orcoles}, \citenamefont
  {Smolin}, \citenamefont {Merkel}, \citenamefont {Rozen}, \citenamefont
  {Keefe}, \citenamefont {Rothwell}, \citenamefont {Ketchen},\ and\
  \citenamefont {Steffen}}]{ka:212-rigettisteffen-prb-trasmonshapphire}%
  \BibitemOpen
  \bibfield  {author} {\bibinfo {author} {\bibfnamefont {C.}~\bibnamefont
  {Rigetti}}, \bibinfo {author} {\bibfnamefont {J.~M.}\ \bibnamefont
  {Gambetta}}, \bibinfo {author} {\bibfnamefont {S.}~\bibnamefont {Poletto}},
  \bibinfo {author} {\bibfnamefont {B.~L.~T.}\ \bibnamefont {Plourde}},
  \bibinfo {author} {\bibfnamefont {J.~M.}\ \bibnamefont {Chow}}, \bibinfo
  {author} {\bibfnamefont {A.~D.}\ \bibnamefont {C\'orcoles}}, \bibinfo
  {author} {\bibfnamefont {J.~A.}\ \bibnamefont {Smolin}}, \bibinfo {author}
  {\bibfnamefont {S.~T.}\ \bibnamefont {Merkel}}, \bibinfo {author}
  {\bibfnamefont {J.~R.}\ \bibnamefont {Rozen}}, \bibinfo {author}
  {\bibfnamefont {G.~A.}\ \bibnamefont {Keefe}}, \bibinfo {author}
  {\bibfnamefont {M.~B.}\ \bibnamefont {Rothwell}}, \bibinfo {author}
  {\bibfnamefont {M.~B.}\ \bibnamefont {Ketchen}}, \ and\ \bibinfo {author}
  {\bibfnamefont {M.}~\bibnamefont {Steffen}},\ }\href {\doibase
  10.1103/PhysRevB.86.100506} {\bibfield  {journal} {\bibinfo  {journal} {Phys.
  Rev. B}\ }\textbf {\bibinfo {volume} {86}},\ \bibinfo {pages} {100506}
  (\bibinfo {year} {2012})}\BibitemShut {NoStop}%
\bibitem [{\citenamefont {Falci}\ \emph {et~al.}(2005)\citenamefont {Falci},
  \citenamefont {D'Arrigo}, \citenamefont {Mastellone},\ and\ \citenamefont
  {Paladino}}]{ka:205-falci-prl}%
  \BibitemOpen
  \bibfield  {author} {\bibinfo {author} {\bibfnamefont {G.}~\bibnamefont
  {Falci}}, \bibinfo {author} {\bibfnamefont {A.}~\bibnamefont {D'Arrigo}},
  \bibinfo {author} {\bibfnamefont {A.}~\bibnamefont {Mastellone}}, \ and\
  \bibinfo {author} {\bibfnamefont {E.}~\bibnamefont {Paladino}},\ }\href
  {\doibase 10.1103/PhysRevLett.94.167002} {\bibfield  {journal} {\bibinfo
  {journal} {Phys. Rev. Lett.}\ }\textbf {\bibinfo {volume} {94}},\ \bibinfo
  {pages} {167002} (\bibinfo {year} {2005})}\BibitemShut {NoStop}%
\bibitem [{\citenamefont {Paladino}\ \emph {et~al.}(2014)\citenamefont
  {Paladino}, \citenamefont {Galperin}, \citenamefont {Falci},\ and\
  \citenamefont {Altshuler}}]{kr:214-paladino-rmp}%
  \BibitemOpen
  \bibfield  {author} {\bibinfo {author} {\bibfnamefont {E.}~\bibnamefont
  {Paladino}}, \bibinfo {author} {\bibfnamefont {Y.}~\bibnamefont {Galperin}},
  \bibinfo {author} {\bibfnamefont {G.}~\bibnamefont {Falci}}, \ and\ \bibinfo
  {author} {\bibfnamefont {B.}~\bibnamefont {Altshuler}},\ }\href {\doibase
  10.1103/RevModPhys.86.361} {\bibfield  {journal} {\bibinfo  {journal} {Rev.
  Mod. Phys.}\ }\textbf {\bibinfo {volume} {86}},\ \bibinfo {pages} {361}
  (\bibinfo {year} {2014})}\BibitemShut {NoStop}%
\bibitem [{\citenamefont {Falci}\ \emph {et~al.}(2013)\citenamefont {Falci},
  \citenamefont {La~Cognata}, \citenamefont {Berritta}, \citenamefont
  {D'Arrigo}, \citenamefont {Paladino},\ and\ \citenamefont
  {Spagnolo}}]{ka:213-falci-prb-stirapcpb}%
  \BibitemOpen
  \bibfield  {author} {\bibinfo {author} {\bibfnamefont {G.}~\bibnamefont
  {Falci}}, \bibinfo {author} {\bibfnamefont {A.}~\bibnamefont {La~Cognata}},
  \bibinfo {author} {\bibfnamefont {M.}~\bibnamefont {Berritta}}, \bibinfo
  {author} {\bibfnamefont {A.}~\bibnamefont {D'Arrigo}}, \bibinfo {author}
  {\bibfnamefont {E.}~\bibnamefont {Paladino}}, \ and\ \bibinfo {author}
  {\bibfnamefont {B.}~\bibnamefont {Spagnolo}},\ }\href {\doibase
  10.1103/PhysRevB.87.214515} {\bibfield  {journal} {\bibinfo  {journal} {Phys.
  Rev. B}\ }\textbf {\bibinfo {volume} {87}},\ \bibinfo {pages} {214515}
  (\bibinfo {year} {2013})}\BibitemShut {NoStop}%
\bibitem [{\citenamefont {Geva}\ \emph {et~al.}(1995)\citenamefont {Geva},
  \citenamefont {Kosloff},\ and\ \citenamefont
  {Skinner}}]{ka:195-geva-jorchemphys-gmerabi}%
  \BibitemOpen
  \bibfield  {author} {\bibinfo {author} {\bibfnamefont {E.}~\bibnamefont
  {Geva}}, \bibinfo {author} {\bibfnamefont {R.}~\bibnamefont {Kosloff}}, \
  and\ \bibinfo {author} {\bibfnamefont {J.~L.}\ \bibnamefont {Skinner}},\
  }\href {\doibase 10.1063/1.468844} {\bibfield  {journal} {\bibinfo  {journal}
  {J. Chem. Phys.}\ }\textbf {\bibinfo {volume} {102}},\ \bibinfo {pages}
  {8541} (\bibinfo {year} {1995})}\BibitemShut {NoStop}%
\bibitem [{\citenamefont {Vitanov}\ \emph {et~al.}(2001)\citenamefont
  {Vitanov}, \citenamefont {Fleischhauer}, \citenamefont {Shore},\ and\
  \citenamefont {Bergmann}}]{kr:201-vitanov-advatmolopt}%
  \BibitemOpen
  \bibfield  {author} {\bibinfo {author} {\bibfnamefont {N.}~\bibnamefont
  {Vitanov}}, \bibinfo {author} {\bibfnamefont {M.}~\bibnamefont
  {Fleischhauer}}, \bibinfo {author} {\bibfnamefont {B.}~\bibnamefont {Shore}},
  \ and\ \bibinfo {author} {\bibfnamefont {K.}~\bibnamefont {Bergmann}},\
  }\href {\doibase DOI: 10.1016/S1049-250X(01)80063-X} {\bibfield  {journal}
  {\bibinfo  {journal} {Adv. in At. Mol. and Opt. Phys.}\ }\textbf {\bibinfo
  {volume} {46}},\ \bibinfo {pages} {55} (\bibinfo {year} {2001})}\BibitemShut
  {NoStop}%
\end{thebibliography}
%
\end{document}